\pdfoutput=1
\documentclass{article}
\usepackage[english]{babel}
\usepackage[letterpaper,top=2cm,bottom=2cm,left=2cm,right=2cm,marginparwidth=1cm]{geometry}
\usepackage{authblk}
\usepackage{amsmath}
\usepackage{amssymb}
\usepackage{amsthm}
\usepackage{bm}
\usepackage{graphicx}
\usepackage{caption}
\usepackage{array}
\usepackage{booktabs}
\usepackage{longtable}
\usepackage[numbers,sort]{natbib}
\usepackage[colorlinks=true,allcolors=blue]{hyperref}
\theoremstyle{definition}
\newtheorem{definition}{Definition}

\title{Adaptive Policy-Guided Error Mitigation (APGEM): A Unified Learning-Based Framework for Noise Resilience in Quantum Machine Learning}
\author[]{Bisma Majid}
\author[]{Shabir Ahmad Sofi}
\author[]{Mir Mohammad Yousuf\thanks{Corresponding author. yousuf\_2022phaite006@nitsri.ac.in}}
\affil{Department of Information Technology\\ National Institute of Technology Srinagar\\ Jammu \& Kashmir, India}
\date{}

\begin{document}
\maketitle

\begin{abstract}
Error mitigation stands between variational quantum algorithms and useful results on today's noisy hardware, yet in practice a mitigation method is usually chosen once, by hand, and applied unchanged for the life of an experiment. This is a poor fit for quantum machine learning (QML), where the circuit and the noise drift as training proceeds, so the best strategy in one regime is often wrong in another. We present a generalized form of Adaptive Policy-Guided Error Mitigation (APGEM), which treats the choice of mitigation technique as an online decision made from an algorithm's own feedback. A single contextual-bandit controller observes signals any variational algorithm already produces, such as fidelity, entropy, reward or loss, and gradient behavior, and selects among Zero-Noise Extrapolation, Probabilistic Error Cancellation, Measurement Error Mitigation, Dynamical Decoupling, and circuit optimization per step. Because it depends only on this shared feedback interface, the framework applies unchanged across supervised, unsupervised, reinforcement, and hybrid quantum-classical learning. Evaluated on five paradigms across 80 algorithm-and-noise conditions, adaptive selection reaches $95.5\%$ of an oracle that always picks the best technique and significantly outperforms all six fixed strategies (Friedman $\chi^{2}=92.18$, $p\approx1.1\times10^{-17}$; pairwise Wilcoxon $p<4\times10^{-6}$). The learned policy is noise-aware: its favored technique shifts as depolarizing and readout error rise. On noise models from eight real IBM devices, APGEM tracks the oracle almost exactly ($99.9\%$ of oracle utility) at a flat cost of roughly $1.3\times10^{4}$ shots from three to eight qubits.
\end{abstract}


\section{Introduction}
\label{sec:intro}
Quantum computing is increasingly recognized as a transformative paradigm capable of addressing problems that remain intractable for classical systems. Within this paradigm, \textit{Quantum Machine Learning (QML)} stands out as a powerful fusion of quantum computation and learning theory, aiming to accelerate pattern recognition, optimization, and generative modeling through quantum-enhanced representations~\cite{schuld2021ml,biamonte2017qml}. However, the current generation of devices, commonly referred to as \textit{Noisy Intermediate-Scale Quantum (NISQ)} hardware, remains fundamentally constrained by decoherence, gate infidelity, and readout errors~\cite{preskill2018nisq,kandala2019em, yousuf2026systematic}. These imperfections distort measurement outcomes, hinder gradient-based training, and destabilize the learning process. As a result, \textit{noise resilience} has become a foundational prerequisite for realizing the practical potential of QML.

To address this challenge, a variety of \textit{error mitigation techniques} have been developed, offering noise suppression without the resource overhead of full quantum error correction. Methods such as \textit{Zero-Noise Extrapolation (ZNE)} and \textit{Probabilistic Error Cancellation (PEC)}~\cite{temme2017em}, \textit{Measurement Error Mitigation (MEM)}~\cite{endo2021hybrid}, \textit{Dynamical Decoupling (DD)}~\cite{viola1999dd}, and circuit-level optimization~\cite{kandala2019em} have each achieved notable success in improving output quality across simulated and experimental platforms. Yet these techniques share a defining characteristic: each is effective only within the regime it was designed for. ZNE performs well when the dominant error is stochastic and moderate; PEC repays its sampling overhead only when accuracy is at a premium; MEM matters when readout confusion dominates; DD earns its place against coherent, rather than stochastic, noise. No single technique is best everywhere, and the differences between them are not marginal. In conventional practice, however, one technique is selected once, by hand, and applied uniformly for the life of an experiment. Mitigation is treated as a \textit{static}, \textit{context-insensitive} preprocessing step, in spite of the fact that real devices exhibit \textit{temporal noise drift}, \textit{circuit-dependent decoherence}, and, in a learning setting, \textit{evolving training dynamics}.

This misalignment is particularly acute in QML, where the quantum model evolves iteratively through gradient updates or reinforcement feedback. The circuit reshapes as parameters move, the effective noise changes with depth and device drift, and the quantity that matters shifts as optimization proceeds, from coarse exploration early in training to fine refinement near convergence. A mitigation strategy that is optimal in the first phase of training may be wasteful, or even harmful, in the last. A static choice cannot track these changes, and this observation motivated the development of \textit{Adaptive Policy-Guided Error Mitigation (APGEM)}, a feedback-driven framework that adjusts its mitigation behavior in response to the state of the computation. Initially formulated for \textit{Quantum Reinforcement Learning (QRL)}, APGEM coupled mitigation to the reinforcement-learning loop, selecting how to correct noisy execution from real-time performance signals rather than fixing a strategy in advance. In doing so, it reframed mitigation from a static post-processing correction into an \textit{active, co-evolving component} of the learning process, demonstrating that feedback-driven adaptation can serve as an intelligent noise controller aligned with task-specific objectives.

The original APGEM formulation, however, was tailored to reinforcement learning and its reward-based feedback. The question that naturally arises is whether APGEM's core principle, \textit{feedback-driven adaptivity in the choice of mitigation}, can be abstracted and generalized to other forms of quantum learning. This generalization is important because modern QML spans a broad spectrum of paradigms, including \textit{Variational Quantum Classifiers (VQCs)}~\cite{schuld2021ml} and \textit{Quantum Neural Networks} for supervised learning, \textit{Quantum Kernel Methods}~\cite{havlicek2019qml} for feature-space embedding, \textit{Variational Quantum Eigensolvers (VQEs)}~\cite{peruzzo2014vqe} for hybrid quantum-classical optimization, and \textit{Quantum Generative Models}~\cite{schuld2019featurehilbert} for probabilistic modeling. Each experiences noise differently and exposes different native performance signals, yet all share a common requirement: preserving learning fidelity under noise. A unifying, model-agnostic approach to adaptive mitigation would therefore represent a significant conceptual and practical advance, building on the variational-algorithm paradigm~\cite{cerezo2021vqa} and offering a scalable and self-optimizing pathway for reliable quantum learning.

In this paper, we propose a \textit{generalized formulation of APGEM} that extends beyond reinforcement learning to a wide range of QML architectures. The central idea is to cast mitigation-technique selection as an \textit{online contextual decision}~\cite{li2010contextual} made from feedback that any variational algorithm already produces. At each decision point, a lightweight controller observes a compact context vector, summarizing output fidelity, measurement entropy, the trend of the task's loss or reward, gradient behavior~\cite{mcclean2018barren}, circuit structure, and the current approximation ratio, and selects one technique from a pool of five (ZNE, PEC, MEM, DD, and circuit optimization), or no mitigation at all. Selection is governed by a \textit{cost-aware utility} that rewards genuine gains in output quality while charging for the sampling overhead each technique incurs, so the controller neither overlooks cheap improvements nor pays for mitigation that does not help. Crucially, the controller depends only on this shared feedback interface and not on the internals of the algorithm it serves, which is what makes a single mechanism applicable across otherwise dissimilar learning paradigms.

To demonstrate this generality, we instantiate the framework on five paradigms spanning the major families of quantum learning:
\begin{itemize}
    \item In \textit{Variational Quantum Classifiers}, the controller stabilizes supervised training by mitigating the noisy measurements from which class predictions are read.
    \item For \textit{Quantum Kernel Methods (QSVMs)}, it preserves kernel fidelity by adapting mitigation to the noise present when kernel entries are estimated.
    \item In \textit{Variational Quantum Eigensolvers}, it protects the energy estimates that drive the hybrid optimization.
    \item In \textit{Quantum Generative Models}, it maintains distributional fidelity of the sampled output as noise varies.
    \item In \textit{Quantum Reinforcement Learning}, the original setting, it recovers reward-aligned policy behavior under noise.
\end{itemize}
Across these paradigms, APGEM functions as a \textit{meta-learner}, a supervisory layer that learns how to mitigate noise by leveraging feedback intrinsic to each learning process, without knowledge of what that process is trying to learn.

The key contributions of this paper are as follows:
\begin{enumerate}
    \item We present a \textbf{generalized, model-agnostic formulation} of Adaptive Policy-Guided Error Mitigation as a feedback-driven contextual-decision mechanism, applicable to any noise-affected variational quantum learning model through a single shared interface.
    \item We provide a \textbf{comprehensive empirical evaluation} across five QML paradigms and 80 algorithm-and-noise conditions, in which adaptive selection reaches $95.5\%$ of an oracle that always chooses the best technique and significantly outperforms every one of six fixed strategies (Friedman $\chi^{2}=92.18$, $p\approx1.1\times10^{-17}$; pairwise Wilcoxon $p<4\times10^{-6}$). We further show that the learned policy is genuinely noise-aware, with its preferred technique shifting as depolarizing and readout error grow.
    \item We \textbf{validate the framework on hardware-calibrated noise} derived from eight real IBM device snapshots, where the controller tracks the oracle almost exactly ($99.9\%$ of oracle utility), demonstrate that its per-decision sampling cost remains constant from three to eight qubits, and provide a regret analysis that attributes most of the residual gap to the irreducible noise floor of finite sampling rather than to the policy.
\end{enumerate}

Through this generalization, APGEM evolves from an application-specific mitigation strategy into a \textit{unified, feedback-oriented principle} for adaptive quantum noise control. This reframes error mitigation not as a static correction layer but as an \textit{interactive, self-optimizing process} that co-evolves with the quantum learner, bridging machine-learning feedback and quantum noise suppression and laying methodological groundwork for noise-resilient quantum learning on near-term hardware.

The remainder of this paper is structured as follows. Section~\ref{related} reviews related work in quantum error mitigation and learning-based noise control. Section~\ref{problem} formalizes the generalized APGEM framework, including its feedback interface, cost-aware utility, and contextual controller. Section~\ref{methodology} details the mitigation pool and the five paradigm instantiations. Section~\ref{exp} describes the noise models, operating conditions, and evaluation protocol. Section~\ref{sec:results} reports the experimental findings, Section~\ref{sec:discussion} discusses their implications and limitations, and Section~\ref{sec:conclusion} concludes and summarizes the main contributions.

\section{Related Work}\label{related}
Noise is still the main barrier to useful computation on near-term, noisy intermediate-scale quantum (NISQ) devices. Full fault-tolerant correction is out of reach at current qubit counts, so quantum error mitigation (QEM) has become the practical alternative: suppress the bias that noise introduces into expectation values, without paying for extra physical qubits. The line of work begins with \cite{Temme_2017}, who introduced mitigation for short-depth circuits and, in doing so, set out the two ideas that still anchor the field, zero-noise extrapolation (ZNE) and probabilistic error cancellation (PEC). Both correct bias through classical post-processing and controlled noise scaling rather than encoding, and both pay for it, in sampling overhead and in sensitivity to noise that drifts or fails to scale cleanly.

The response to those limits was to stop modelling the noise explicitly and start learning it. \cite{strikis2021learning} do this with near-Clifford circuits, which are classically simulable and therefore give ground-truth outputs for free. Pairing those with the corresponding noisy runs yields training data for a learned mitigation map that needs no device calibration. The method performed well on IBM hardware, though it carries the usual costs of a learned model: a tendency to overfit, and the expense of generating enough varied training circuits.

\cite{czarnik2021error} sharpen the same idea into Clifford Data Regression (CDR), fitting a linear model on pairs of noisy and classically computed expectation values from near-Clifford circuits and using it to correct the observables of harder, non-Clifford circuits. CDR cuts bias substantially and makes the trade-off between accuracy, sampling cost, and circuit depth explicit. \cite{Lowe_2021} then fold ZNE and CDR into a single method, variable-noise Clifford data regression (vnCDR), training on near-Clifford circuits run at several amplified noise levels so that extrapolation and regression reinforce each other. On an eight-qubit Ising energy estimate vnCDR reduced the absolute error by a factor of $33$ over the unmitigated result, and by factors of $20$ and $1.8$ over ZNE and CDR used alone.

These methods all assume the noise sits still. \cite{henao2023adaptive} drop that assumption with Adaptive KIK, which retunes its mitigation on the fly from real-time noise estimates and so copes with correlated, time-varying noise. It is the closest prior work in spirit to ours, treating mitigation as a feedback loop rather than a fixed recipe.

Two broader results frame the space. \cite{cai2023quantum} survey QEM by resource cost, scalability, and applicability, and single out quantum machine learning (QML) as a setting where mitigation remains poorly integrated, since noise corrupts the gradient estimates that training depends on. \cite{takagi2022fundamental} give the hard limit: exact cancellation can demand a sample count that grows exponentially with circuit depth, which is why frugal, approximate mitigation is the only realistic option at scale. On the practical side, refinements such as the identity-insertion scheme of \cite{he2020zero} have made ZNE more stable and easier to run on hardware.

Read together, the trajectory is clear: from analytical suppression (ZNE, PEC), to data-driven regression, to adaptive control. What is missing is the last step for machine learning. Almost every method above is static and tied to a single circuit, whereas QML training moves the circuit and the noise at the same time, and no fixed strategy stays optimal across that trajectory. This is the gap the Adaptive Policy-Guided Error Mitigation (APGEM) framework and its generalization target: one learning-based controller that selects and adapts mitigation as training proceeds, rather than committing to a technique in advance.

\begin{longtable}{p{0.5cm} p{3.5cm} p{3cm} p{3cm} p{2.5cm} p{3.5cm}}
\caption{Summary of key literature on quantum error mitigation (QEM) relevant to Adaptive Policy-Guided Error Mitigation (APGEM).}
\label{tab:QEM_literature_summary} \\
\hline
\textbf{Ref} & \textbf{Key Contribution} & \textbf{Method Used} & \textbf{Target Problem} & \textbf{Noises Addressed} & \textbf{Error Mitigation Used and Focus} \\
\hline
\endfirsthead
\multicolumn{6}{c}{{\tablename} \thetable{} -- continued from previous page} \\
\hline
\textbf{Ref} & \textbf{Key Contribution} & \textbf{Method Used} & \textbf{Target Problem} & \textbf{Noises Addressed} & \textbf{Error Mitigation Used and Focus} \\
\hline
\endhead
\hline \multicolumn{6}{r}{{Continued on next page}} \\
\endfoot
\hline
\endlastfoot
\cite{liao2024machine} & Systematic ML-QEM study showing classical models can emulate conventional mitigation at lower runtime overhead. & Supervised models (linear regression, random forests, MLPs, graph neural networks) predicting mitigated observables. & Scalable, low-overhead mitigation across diverse circuits and device-noise profiles. & Device-level stochastic noise (interpolation and extrapolation). & ML-QEM; reduces sampling overhead by mimicking traditional QEM. \\
\cite{muqeet2024machine} & ML-based mitigation framed as a reliable quantum software-development workflow on IBM hardware. & Machine-learning correction of measured outputs, evaluated as a software-engineering pipeline. & Dependable execution of quantum programs under device noise. & Hardware noise on IBM superconducting devices. & Learning-based QEM from a software-reliability perspective. \\
\cite{adeniyi2025adaptive} & Adaptive neural network that dynamically adjusts expectation values from error characteristics on a 127-qubit device. & Neural network mapping noisy to corrected observables, adapting to device error profiles. & Dynamic, device-aware mitigation on large superconducting processors. & Time- and device-dependent hardware noise. & Adaptive learning-based QEM; neural analogue to policy-guided adaptation. \\
\cite{korolev2026machine} & Practical ML-QEM for variational algorithms with per-task model selection and transfer across Hamiltonians. & Regression models (Ridge, XGBoost) trained on simulated near-Clifford data and selected per task. & Parameter- and Hamiltonian-transferable mitigation for VQAs, benchmarked on VQE. & Circuit- and depth-dependent noise, high-noise regime. & ML-QEM with model selection; conceptually aligned with adaptive selection. \\
\cite{Temme_2017} & Foundational framework introducing zero-noise extrapolation (ZNE) and probabilistic error cancellation (PEC) for NISQ devices. & Analytical extrapolation and quasi-probability correction. & Bias reduction in short-depth quantum circuits. & Gate and readout noise (Markovian). & ZNE and PEC; analytical and hardware-compatible, but sampling-intensive. \\
\cite{strikis2021learning} & Learning-based QEM trained on near-Clifford circuits, removing the need for an explicit noise model. & Data-driven regression on noisy--ideal circuit pairs. & Hardware noise suppression without explicit noise models. & Device-specific stochastic and coherent noise. & Learning-based QEM; emphasises adaptability and noise-model independence. \\
\cite{czarnik2021error} & Clifford Data Regression (CDR) for data-driven bias correction using classically simulable proxies. & Linear regression on Clifford-circuit datasets. & Estimating ideal observables for non-Clifford circuits. & Hardware-dependent stochastic noise. & Regression-based QEM; balances accuracy, sampling cost, and scalability. \\
\cite{Lowe_2021} & Variable-noise Clifford Data Regression (vnCDR) unifying ZNE and CDR into a single estimator. & Regression on near-Clifford circuits evaluated at multiple amplified noise levels. & Robust observable estimation across noise regimes and circuit depths. & Hardware and depth-dependent stochastic noise. & Hybrid data-driven QEM; combines extrapolation and regression for robustness. \\
\cite{henao2023adaptive} & Adaptive KIK for dynamic, pulse-based inverse-evolution error mitigation. & Adaptive inverse evolution driven by real-time noise estimation. & Circuit robustness under time-dependent noise. & Correlated and temporally varying noise. & Adaptive mitigation guided by device feedback; enables real-time policy adjustment. \\
\cite{cai2023quantum} & Comprehensive review and taxonomy of QEM methods, evaluating trade-offs and scalability. & Comparative survey of QEM techniques. & Classifying QEM methods and identifying open gaps. & General device and circuit-level noise. & Summarises mitigation trade-offs; flags QML integration as an open challenge. \\
\cite{takagi2022fundamental} & Establishes the fundamental sampling limits of quantum error mitigation. & Information-theoretic and sampling-complexity analysis. & Feasibility of unbiased mitigation at scale. & General (arbitrary) noise channels. & Defines theoretical sampling limits and frugality boundaries for QEM. \\
\cite{he2020zero} & Enhanced ZNE via identity insertion for more stable noise scaling. & Refined noise-scaling and extrapolation. & Practical, stable implementation of ZNE on hardware. & Gate and measurement noise. & Optimised extrapolation for greater noise-scaling consistency. \\
\hline
\end{longtable}

\section{Problem Formulation}
\label{problem}
\begin{figure}[h!]
\centering
\includegraphics[width=\linewidth]{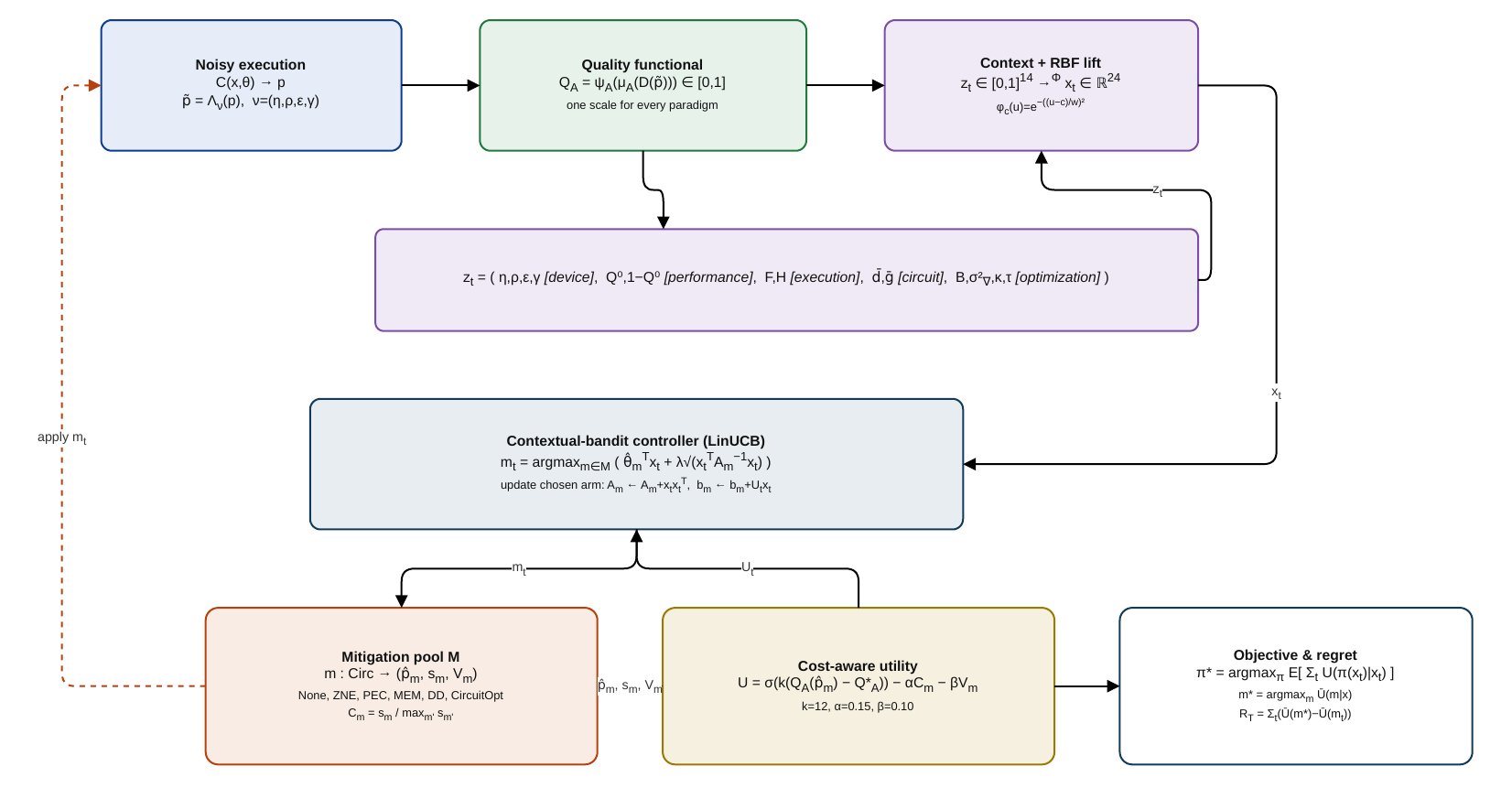}
\caption{The generalized APGEM problem as an online decision loop. In the \emph{perception} stage, a noisy execution of the variational circuit is reduced to a normalized quality functional $Q_{\mathcal{A}}$ (Definition~\ref{def:quality}) and a feedback context $\bm z_t$ that is lifted to $\bm x_t\in\mathbb{R}^{24}$ by a radial-basis map (Definition~\ref{def:context}, Eq.~\ref{eq:rbf}). In the \emph{decision} stage, a contextual-bandit controller (Eq.~\ref{eq:linucb}) selects a technique $m_t$ from the mitigation pool (Definition~\ref{def:pool}), which is scored by the cost-aware utility $U$ (Definition~\ref{def:utility}, Eq.~\ref{eq:utility}) and used both to update the controller and to define the objective and regret against the expected-utility oracle (Definition~\ref{def:oracle}). The controller reads only the feedback interface, never the algorithm's internals, which is what makes a single mechanism applicable across all five paradigms.}
\label{fig:problem}
\end{figure}

This section formalizes adaptive mitigation as an online decision problem. We first fix the setting of a variational quantum learning algorithm running under noise (Section~\ref{sec:pf-setting}), then define the three objects that let a single controller serve any such algorithm: a normalized \emph{quality functional} that places every paradigm's native performance measure on a common scale (Section~\ref{sec:pf-quality}), an \emph{operating context} assembled from feedback the algorithm already produces (Section~\ref{sec:pf-context}), and a \emph{mitigation pool} of cost-bearing correction maps (Section~\ref{sec:pf-pool}). A cost-aware utility (Section~\ref{sec:pf-utility}) then turns technique selection into a contextual bandit (Section~\ref{sec:pf-bandit}), whose performance we measure against an expected-utility oracle (Section~\ref{sec:pf-oracle}). Figure~\ref{fig:problem} summarizes the resulting loop and fixes the notation used throughout.

\subsection{Setting}
\label{sec:pf-setting}
We consider a variational quantum learning algorithm $\mathcal{A}=(\mathcal{C},\Theta,\mathcal{D},\mu_{\mathcal{A}})$, where $\mathcal{C}:\mathcal{X}\times\Theta\to\mathsf{Circ}$ maps an input $x\in\mathcal{X}$ and trainable parameters $\theta\in\Theta\subseteq\mathbb{R}^{d}$ to a parameterized circuit, $\mathcal{D}$ is a decoder that maps a measured outcome distribution to the algorithm's output, and $\mu_{\mathcal{A}}$ is the algorithm's native performance indicator, such as classification accuracy, kernel-target accuracy, estimated energy, a distributional distance, or an expected return. This tuple abstracts every paradigm we study, and it is the only interface the mitigation layer is allowed to see.

Executed ideally, the circuit $\mathcal{C}(x,\theta)$ produces an outcome distribution $p$. On NISQ hardware~\cite{preskill2018nisq}, execution instead realizes a completely positive trace-preserving (CPTP) map $\Lambda_{\nu}$ that depends on the device's \emph{operating condition} $\nu$, so the observed distribution is
\begin{equation}
\tilde{p} \;=\; \Lambda_{\nu}(p).
\end{equation}
We summarize the operating condition by $\nu=(\eta,\rho,\varepsilon,\gamma)$, collecting the depolarizing rate $\eta$, the readout error $\rho$, a coherent idle-rotation angle $\varepsilon$, and an amplitude-damping rate $\gamma$. The distinction among these components matters because different mitigation techniques target different terms: measurement mitigation acts on $\rho$, dynamical decoupling on $\varepsilon$, and extrapolation-based methods on the stochastic part of $\eta$~\cite{temme2017em,viola1999dd}. The condition $\nu$ is not static; it drifts with device calibration and changes with circuit depth as training proceeds.

\subsection{Quality functional}
\label{sec:pf-quality}
A single controller can serve heterogeneous algorithms only if their dissimilar performance measures are expressed on one comparable scale. We therefore normalize each algorithm's native indicator into a common quality signal.

\begin{definition}[Quality functional]
\label{def:quality}
For an algorithm $\mathcal{A}$ we fix a monotone map $\psi_{\mathcal{A}}:\operatorname{range}(\mu_{\mathcal{A}})\to[0,1]$ and define the quality functional
\begin{equation}
Q_{\mathcal{A}}(\tilde{p}) \;=\; \psi_{\mathcal{A}}\!\big(\mu_{\mathcal{A}}(\mathcal{D}(\tilde{p}))\big)\in[0,1],
\end{equation}
oriented so that larger values are better.
\end{definition}

By construction $Q_{\mathcal{A}}$ discards everything algorithm-specific and retains only ``how well is the computation doing, on a $0$ to $1$ scale.'' A classifier reports accuracy directly, an eigensolver maps its energy error onto $[0,1]$ against the spectral width, a generative model reports one minus a total-variation distance to its target, and a reinforcement-learning agent reports a normalized return. The controller sees only $Q_{\mathcal{A}}$ and never the definition behind it, which is precisely what lets one mechanism apply across supervised, unsupervised, reinforcement, and hybrid learning. The concrete instantiations of $\psi_{\mathcal{A}}$ for the five paradigms are given in Section~\ref{methodology}.

\subsection{Operating context}
\label{sec:pf-context}
At each decision step $t$ the controller must choose a mitigation technique before it can know the outcome. It does so from a \emph{context vector} built entirely from signals a variational pipeline already computes, together with a cheap unmitigated probe of the current circuit.

\begin{definition}[Context]
\label{def:context}
The raw context at step $t$ is
\begin{equation}
z_t = \big(\;
\underbrace{\eta_t,\rho_t,\varepsilon_t,\gamma_t}_{\text{device}},\;
\underbrace{Q^{0}_t,\,1-Q^{0}_t}_{\text{performance}},\;
\underbrace{F_t,\,H_t}_{\text{execution}},\;
\underbrace{\bar d_t,\,\bar g_t}_{\text{circuit}},\;
\underbrace{B_t,\,\sigma^{2}_{\nabla,t},\,\kappa_t,\,\tau_t}_{\text{optimization}}
\;\big)\in[0,1]^{14},
\end{equation}
where $Q^{0}_t$ is the quality of an unmitigated probe and $1-Q^{0}_t$ its deficit; $F_t$ is the distribution fidelity of the probe to its ideal reference and $H_t$ its normalized entropy; $\bar d_t$ and $\bar g_t$ are the normalized circuit depth and two-qubit gate count; and $B_t$, $\sigma^{2}_{\nabla,t}$, $\kappa_t$, and $\tau_t$ are the remaining shot budget, the gradient variance~\cite{mcclean2018barren}, a convergence-rate estimate, and the training progress. All entries are normalized to $[0,1]$.
\end{definition}

The optimal technique is not a monotone function of the noise rates: no mitigation is best at very low noise, cancellation-based methods dominate at moderate noise, and beyond a point mitigation cannot recover usable quality and cheap execution is again preferable. A model that is linear in $z_t$ cannot separate these regimes, so we lift the noise axes through a radial-basis expansion,
\begin{equation}
x_t = \Phi(z_t) = \Big(z_t,\;\{\phi_c(\eta_t)\}_{c\in\mathcal{K}_\eta},\;
\{\phi_c(\rho_t)\}_{c\in\mathcal{K}_\rho},\;
\{\phi_c(\varepsilon_t)\}_{c\in\mathcal{K}_\varepsilon}\Big),
\qquad
\phi_c(u)=\exp\!\Big(-\big((u-c)/w\big)^{2}\Big),
\label{eq:rbf}
\end{equation}
with six centers on the depolarizing axis and two each on the readout and coherent axes. This yields a lifted context $x_t\in\mathbb{R}^{24}$, in which each technique's region of optimality becomes approximately linearly separable. The lift is the mechanism that makes a linear controller sufficient across regimes.

\subsection{Mitigation pool}
\label{sec:pf-pool}
The available corrections form a finite pool of maps that share one signature, so the controller can treat them interchangeably.

\begin{definition}[Mitigation pool]
\label{def:pool}
The pool is a finite set $\mathcal{M}$ in which each technique $m\in\mathcal{M}$ is a map
\begin{equation}
m:\mathsf{Circ}\longrightarrow(\hat p_m,\,s_m,\,V_m),
\end{equation}
returning a mitigated distribution $\hat p_m$, a sampling cost $s_m$ measured in shots, and an estimator variance $V_m$. We use
\begin{equation}
\mathcal{M}=\{\textsc{None},\textsc{ZNE},\textsc{PEC},\textsc{MEM},\textsc{DD},\textsc{CircuitOpt}\},
\end{equation}
spanning four error classes: stochastic gate noise (ZNE, PEC), measurement error (MEM), coherent idle error (DD), and structural noise reduced by compilation (CircuitOpt), together with the null option of no mitigation~\cite{temme2017em,giurgicatiron2020zne,vandenberg2023pec,viola1999dd}. The normalized cost of a technique is $C_m = s_m / \max_{m'\in\mathcal{M}} s_{m'} \in (0,1]$.
\end{definition}

The pool is open under this signature: any correction that maps a circuit to a $(\hat p, s, V)$ triple can be added without modifying the controller, and the per-paradigm implementations are deferred to Section~\ref{methodology}.

\subsection{Cost-aware utility}
\label{sec:pf-utility}
A controller that maximized quality alone would always select the most powerful correction, ignoring that some techniques cost several times as many shots as others. We therefore score a technique by a utility that pays for quality but charges for cost and instability.

\begin{definition}[Cost-aware utility]
\label{def:utility}
Given a task-specific usability threshold $Q^{\ast}_{\mathcal{A}}$, the utility of applying technique $m$ in context $x_t$ is
\begin{equation}
U(m\mid x_t)
= \sigma\!\Big(k\big(Q_{\mathcal{A}}(\hat p_m)-Q^{\ast}_{\mathcal{A}}\big)\Big)
- \alpha\,C_m - \beta\,V_m,
\qquad \sigma(u)=\frac{1}{1+e^{-u}},
\label{eq:utility}
\end{equation}
with sharpness $k=12$ and cost and variance weights $\alpha=0.15$, $\beta=0.10$.
\end{definition}

Centering the logistic at a usability threshold, rather than at the unmitigated quality, encodes a fact a linear reward cannot: a quality gain that carries the algorithm \emph{across} the usable region is worth far more than an equal gain in an already-saturated one. As a consequence, an expensive technique is rational only where it changes the outcome, and near saturation the sigmoid flattens so the cost term dominates and the controller correctly falls back to cheap execution. The thresholds $Q^{\ast}_{\mathcal{A}}$ are set per paradigm and reported in Section~\ref{exp}.

\subsection{Adaptive mitigation as a contextual bandit}
\label{sec:pf-bandit}
At each step the controller observes $x_t$, selects one technique, applies it, and observes only the utility of the technique it chose; this is the decision stage of Figure~\ref{fig:problem}. It never sees the utility of the alternatives, which rules out supervised or counterfactual updates and places the problem squarely in the contextual-bandit setting~\cite{li2010contextual}.

\begin{definition}[Adaptive mitigation problem]
\label{def:bandit}
Let contexts $x_t$ be drawn from the deployment distribution of operating conditions. A policy $\pi$ maps a context to a technique, $m_t=\pi(x_t)$, and the objective is to maximize the expected cumulative utility
\begin{equation}
\pi^{\ast}=\arg\max_{\pi}\;\mathbb{E}\!\left[\sum_{t=1}^{T} U\big(\pi(x_t)\mid x_t\big)\right],
\end{equation}
where the expectation is over the contexts and the shot noise of execution.
\end{definition}

We estimate, per technique, a linear model of the utility in the lifted context and select optimistically. Maintaining $A_m$ and $b_m$ for each arm, the estimate is $\hat\theta_m = A_m^{-1} b_m$, and the selection rule is
\begin{equation}
m_t=\arg\max_{m\in\mathcal{M}}
\Big(\hat\theta_m^{\top} x_t + \lambda\,\sqrt{\,x_t^{\top} A_m^{-1} x_t\,}\Big),
\label{eq:linucb}
\end{equation}
after which only the chosen arm is updated, $A_{m_t}\leftarrow A_{m_t}+x_t x_t^{\top}$ and $b_{m_t}\leftarrow b_{m_t}+U_t\,x_t$, by the Sherman--Morrison identity. The exploration coefficient $\lambda$ is held fixed rather than annealed, so exploration persists wherever a per-arm posterior remains uncertain, which is the behavior needed when the device drifts. For linearly realizable utilities this rule attains the standard $\tilde{\mathcal{O}}(\sqrt{T})$ contextual-bandit regret~\cite{li2010contextual}; we treat that as motivation for the design rather than as a claim about the quantum setting, and we report measured regret against an empirical oracle below.

\subsection{Oracle and regret}
\label{sec:pf-oracle}
To judge the policy we compare it against the best achievable choice under the same information.

\begin{definition}[Expected-utility oracle and regret]
\label{def:oracle}
The oracle policy selects, at each context, the technique of highest \emph{expected} utility,
\begin{equation}
m^{\ast}(x)=\arg\max_{m\in\mathcal{M}}\ \bar U(m\mid x),
\qquad
\bar U(m\mid x)=\mathbb{E}\big[U(m\mid x)\big],
\end{equation}
with the expectation estimated by repeated execution at fixed context. The expected regret of a policy over a horizon $T$ is
\begin{equation}
\mathcal{R}_T=\sum_{t=1}^{T}\Big(\bar U\big(m^{\ast}(x_t)\big)-\bar U(m_t)\Big).
\end{equation}
\end{definition}

We deliberately compare against the \emph{expected}-utility oracle rather than a per-realization one. An oracle that peeks at the sampled outcome and takes the best technique after the fact carries an irreducible advantage from shot noise that no causal policy can match; charging that gap to the controller would report nonzero regret even for an optimal policy. Separating the two, the genuine policy suboptimality from the sampling floor, is what makes the reported gains honest, and we quantify both in Section~\ref{sec:results}.

\section{Methodology}\label{methodology}
The previous section posed adaptive mitigation as an online decision problem. What follows is how we actually built and tested it. The short version is that the mitigation logic lives entirely outside the learning algorithms it serves, talks to them through one narrow interface, and is evaluated on a precomputed table of quantum executions so that every controller we compare sees exactly the same conditions.

The system has three parts. Each learning algorithm is wrapped in a \emph{task adapter} that, when handed a mitigation technique, runs the algorithm under that technique and reports back a small tuple: the normalized quality $Q_{\mathcal{A}}$, the shots consumed, the estimator variance, the distribution fidelity, and the normalized entropy. Alongside it the adapter exposes a gradient variance probe, basic circuit statistics, and its usability threshold $Q^{\ast}_{\mathcal{A}}$. That is the entire contract, and it is the only algorithm-specific code in the system. The \emph{mitigation pool} sits on the other side of the contract, where every technique implements a single method that takes a circuit and returns a corrected distribution together with what it cost. Between them sits the \emph{controller}, which sees the lifted context and the utility of whichever arm it picked, and nothing else. The practical consequence is that adding a sixth paradigm means writing one adapter and touching nothing else, and adding a seventh mitigation technique means writing one class and touching no algorithm. That is a modest claim about software structure rather than about physics, but it is the reason a single controller can serve five very different learners. All quantum execution runs on the Qiskit Aer state-vector simulator under explicit noise models.

\subsection{The five learning paradigms}
\begin{figure}[htbp]
\centering
\includegraphics[width=1\linewidth]{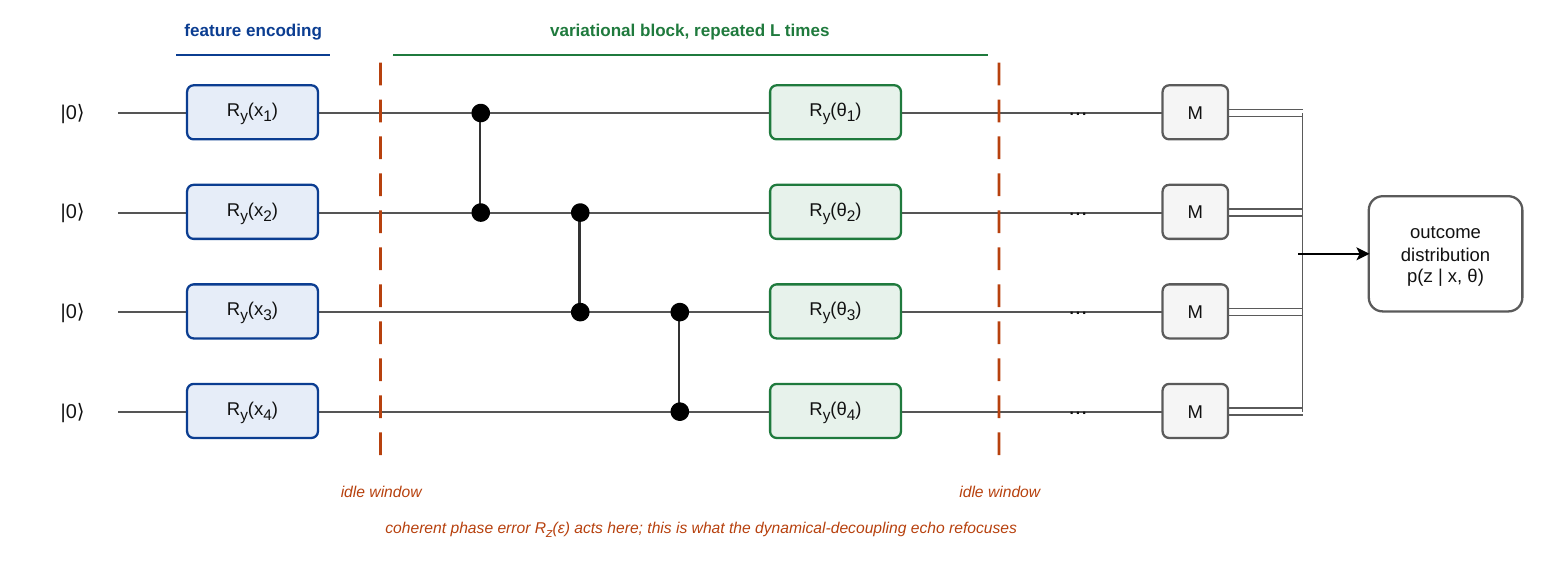}
\caption{The shared ansatz, drawn for $n_q=4$. Every paradigm uses this circuit, so observed differences come from the learning task and the noise rather than from circuit design. The dashed lines are the barriers of Equation~\ref{eq:ansatz}: they mark the idle windows where the coherent phase error acts and, equivalently, the only places a dynamical-decoupling echo has anything to refocus. Removing them would leave that arm of the mitigation pool with no error class of its own.}
\label{fig:ansatz}
\end{figure}

Every adapter is built on the same hardware-efficient ansatz, so that whatever differences we observe come from the learning task and the noise rather than from circuit design. For $n_q$ qubits and $L$ layers the circuit encodes the input in a layer of $R_y$ rotations and then repeats a block of nearest-neighbour controlled-$Z$ gates followed by trainable $R_y$ rotations,
\begin{equation}
U(x,\theta)=\prod_{\ell=1}^{L}
\Big[\textstyle\bigotimes_{q} R_y(\theta_{\ell,q})\Big]
\Big[\textstyle\prod_{q=1}^{n_q-1} \mathrm{CZ}_{q,q+1}\Big]
\ \Big[\textstyle\bigotimes_{q} R_y(x_q)\Big],
\label{eq:ansatz}
\end{equation}
with a barrier closing each block (Figure~\ref{fig:ansatz}). Those barriers matter more than they look. They mark the idle windows where the coherent phase error described below acts, and they are what an echo sequence has to work with. Without them dynamical decoupling would have nothing to refocus and would be a strictly redundant arm. We use $n_q=4$ and $L=2$ throughout unless stated otherwise.

The five adapters differ in what they are trying to do and therefore in what "performing well" means, which is exactly the difficulty the quality functional exists to resolve. Table~\ref{tab:adapters} collects the normalizations. The variational classifier separates two overlapping Gaussian clusters in angle space by the sign of the parity observable, and reports accuracy. The kernel method estimates kernel entries from the all-zeros outcome probability of the overlap circuit $U^{\dagger}(x_j)U(x_i)$, trains a support-vector machine on the resulting noisy kernel, and reports held-out accuracy, so the metric reflects the decision a user would actually make rather than the kernel matrix in isolation. The eigensolver targets a transverse-field Ising chain, reconstructs the energy from $Z$- and $X$-basis measurements, and reports the energy error normalized by the spectral width. The generative model is a Born machine whose noiseless output is taken as the distribution it should reproduce, so that the measured total-variation distance isolates drift caused by noise. The reinforcement learner solves a seven-node routing problem by aggregating the measurement distribution into contiguous blocks over the distance-ordered feasible actions and sampling from them, reporting the ratio of the brute-force optimum to the tour it found.

Two of these deserve a word of justification. For the classifier and the eigensolver we first optimize the variational parameters on the noiseless simulator, so that mitigation is asked to recover the performance of a model that had genuinely converged rather than to flatter one that never worked. Reporting a mitigation gain on an untrained circuit would be close to meaningless. And the reinforcement policy samples its action rather than taking the argmax, because a policy that always takes the modal action is largely insensitive to distortion of the distribution's tail, which would understate the effect of noise in precisely the setting where a practitioner would care about it.

\begin{table}[htbp]
\centering
\caption{The five task adapters. Each maps a different native indicator onto the common quality scale; the controller sees only $Q_{\mathcal{A}}$ and the threshold.}
\label{tab:adapters}
\small
\begin{tabular}{llp{4.2cm}p{3.9cm}c}
\toprule
Adapter & Paradigm & Native indicator & Normalization $\psi_{\mathcal{A}}$ & $Q^{\ast}_{\mathcal{A}}$ \\
\midrule
VQC  & supervised            & parity-readout classification accuracy & identity on $[0,1]$ & 0.75 \\
QSVM & supervised (kernel)   & test accuracy of an SVM on the quantum kernel & identity on $[0,1]$ & 0.75 \\
VQE  & hybrid eigensolver    & energy $E$ of a transverse-field Ising chain & $1-\min\!\big(1,|E-E_0|/|E_{\max}-E_0|\big)$ & 0.80 \\
QGen & unsupervised generative & total-variation distance to the target & $1-\mathrm{TV}$ & 0.70 \\
QRL  & reinforcement         & tour cost $c$ on a seven-node routing MDP & $c^{\ast}/c$, clipped to $[0,1]$ & 0.60 \\
\bottomrule
\end{tabular}
\end{table}

\subsection{What the controller can choose from}
The pool holds six options, chosen to span genuinely different error classes rather than to inflate the number of arms. Doing nothing is one of them, and it is not a placeholder: at low noise it is frequently the right answer, and a framework that cannot decide to spend nothing is not really cost-aware. Zero-noise extrapolation folds the circuit globally as $U\!\to\!U(U^{\dagger}U)^{k}$ at scale factors $\{1,3,5\}$ and extrapolates the outcome distribution back to zero noise. Probabilistic error cancellation inverts an effective global depolarizing channel, $\hat p=(\tilde p-p_{\mathrm{eff}}/2^{n_q})/(1-p_{\mathrm{eff}})$, with $p_{\mathrm{eff}}$ estimated from the circuit's gate count and the device rate, and draws extra shots to absorb the variance that any such inversion amplifies. Measurement mitigation inverts a tensored confusion matrix and is charged for its calibration shots rather than being given them for free. Dynamical decoupling inserts an $X\!-\!X$ echo pair into each idle window, cancelling a deterministic idle phase to first order, which is the one error class none of the stochastic methods can touch~\cite{viola1999dd}. Circuit optimization simply compiles the circuit harder, at optimization level three into an $\{R_z,\sqrt{X},X,\mathrm{CZ}\}$ basis, reducing the noise injected instead of correcting it afterwards. Corrected distributions are renormalized onto the simplex, and each technique's cost is read off the shot counters rather than assumed from theory~\cite{temme2017em,giurgicatiron2020zne,vandenberg2023pec}.

The controller itself is the LinUCB rule of Equation~\ref{eq:linucb}, implemented with the usual care. Each arm keeps $A_m$ initialized to the identity for ridge regularization and $b_m$ carrying a small optimistic offset so that every technique gets tried early rather than being starved by an unlucky first draw. The inverse $A_m^{-1}$ is maintained by Sherman--Morrison, which keeps an update at $O(D^2)$ instead of the $O(D^3)$ a fresh inversion would cost. We hold the exploration coefficient fixed rather than annealing it, because a device that drifts will eventually invalidate whatever the controller learned early, and an annealed schedule would have stopped looking by then. Only the selected arm is updated, so no counterfactual information leaks into learning. At evaluation the controller drops the exploration bonus and acts greedily. For comparison we implement three alternatives behind the identical interface: Bayesian linear Thompson sampling, a context-free $\varepsilon$-greedy selector, and a static controller for each fixed technique.

\subsection{Noise, and how the study is run}
We use two families of noise. The constructed grid crosses depolarizing rates $\{0.01,0.05,0.12,0.18\}$ with readout error $\{0,0.10\}$ and a coherent idle rotation $\{0,0.5\}$, giving sixteen operating conditions. Depolarizing and amplitude-damping errors attach to single-qubit gates, entangling gates carry twice the depolarizing rate to reflect their higher empirical error, and readout error is a symmetric bit-flip confusion matrix. The coherent term is handled separately: it is injected as a deterministic $R_z(\varepsilon)$ at each barrier rather than through the stochastic noise model. That choice is deliberate and worth being explicit about, because a snapshot noise model contains no coherent component at all, and had we not added one the dynamical-decoupling arm would have been included under conditions where it could never win.

The second family replaces all of this with measured hardware. We build noise models directly from the calibration snapshots of eight IBM backends, each carrying its own $T_1$ and $T_2$ relaxation times, per-gate error rates, and per-qubit readout confusion matrices. Across those eight devices the measured readout error spans $0.014$ to $0.037$, two-qubit gate error $0.006$ to $0.013$, and $T_1$ runs from $62$ to $148\,\mu$s. Crossing each device with the coherent term again gives sixteen conditions. This is the closest we get to hardware without running on it, and we are explicit in Section~\ref{sec:results} about what that does and does not establish.

Running every controller against live simulation would be both expensive and unfair, since two controllers would never see the same noise realizations. So we precompute a response table instead: for every combination of task, condition and technique, the adapter is evaluated over repeated independent executions and the resulting records are cached. Every controller is then run against that identical table. This buys three things at once. The comparison becomes paired by construction, the expected-utility oracle of Definition~\ref{def:oracle} can be computed exactly by averaging over the stored repetitions rather than approximated, and the controller experiments become reproducible without re-running the simulator. The unmitigated entry of each cell does double duty as the probe that supplies $Q^{0}_t$, $F_t$ and $H_t$ to the context, which also guarantees the controller never sees the outcome of a technique it did not choose.

For the statistics we treat the task-condition pair as the blocking factor and stay non-parametric throughout. A Friedman omnibus test runs across the seven strategies, the adaptive controller and the six fixed techniques, followed by a Nemenyi post-hoc test with its critical difference for the ranking, and pairwise Wilcoxon signed-rank tests of the controller against each fixed technique, each reported with Cliff's delta as an effect size. Means carry bootstrap $95\%$ confidence intervals. We avoid $t$-tests here on purpose: utilities across conditions are neither independent nor close to normally distributed, and a parametric test would be reporting a confidence it has not earned.

\subsection{What it costs}
It is worth separating the cost of the mitigation layer from the cost of simulating the algorithms underneath it, because only the first is attributable to the framework. Selecting an arm requires one score per technique, each an inner product and a quadratic form in the lifted context, and the update is a rank-one correction, so per decision the controller costs $O(|\mathcal{M}|D^2)$ in both time and memory, with $D=24$ and $|\mathcal{M}|=6$ here. Neither term involves the qubit count or the horizon. The controller does not grow as training proceeds or as the register widens.

The cost that actually bites on hardware is measurement, and Table~\ref{tab:overhead-method} gives what we measured. Dynamical decoupling and circuit optimization are free in shots because they change the circuit rather than the sampling budget. Measurement mitigation costs twice unmitigated execution through its calibration, zero-noise extrapolation three times through its folded evaluations, and probabilistic error cancellation four times through variance-controlled resampling. The important point is that the controller \emph{selects} rather than adds, so its realized cost is whatever mixture its policy induces, bounded below by the cheapest arm and above by the most expensive. This is the mechanism through which the cost-aware utility does its work: an expensive technique is bought only where it changes the outcome, and the framework never incurs a fixed multiplicative penalty the way a statically chosen technique does.

\begin{table}[htbp]
\centering
\caption{Measured sampling overhead per technique, normalized to unmitigated execution and pooled over all paradigms and conditions.}
\label{tab:overhead-method}
\begin{tabular}{lrr}
\toprule
Technique & Shots per application & Overhead \\
\midrule
NoMitigation & 6{,}554  & $1.00\times$ \\
DD           & 6{,}554  & $1.00\times$ \\
CircuitOpt   & 6{,}554  & $1.00\times$ \\
MEM          & 13{,}107 & $2.00\times$ \\
ZNE          & 19{,}661 & $3.00\times$ \\
PEC          & 26{,}214 & $4.00\times$ \\
\bottomrule
\end{tabular}
\end{table}

Scaling behaves the way the complexity argument predicts. Table~\ref{tab:scal-method} sweeps the register from three to eight qubits. Circuit depth grows linearly from seven to seventeen and the parameter count doubles, yet the sampling cost of a complete six-technique decision stays fixed at $13{,}312$ shots and the wall time per decision barely moves. System size is absorbed by the adapter, not by the mitigation layer.

\begin{table}[htbp]
\centering
\caption{Cost of a full mitigation decision as the register grows (VQE adapter). The sampling cost per decision does not depend on the qubit count.}
\label{tab:scal-method}
\begin{tabular}{rrrrr}
\toprule
Qubits & Depth & Parameters & Shots per decision & Wall time (s) \\
\midrule
3 & 7  & 6  & 13{,}312 & 5.1 \\
4 & 9  & 8  & 13{,}312 & 2.2 \\
5 & 11 & 10 & 13{,}312 & 2.2 \\
6 & 13 & 12 & 13{,}312 & 2.4 \\
7 & 15 & 14 & 13{,}312 & 2.6 \\
8 & 17 & 16 & 13{,}312 & 2.5 \\
\bottomrule
\end{tabular}
\end{table}

None of this makes quantum simulation cheap. State-vector simulation of a circuit with $n_q$ qubits, depth $L$ and $D$ gates per layer still costs $O(LD2^{n_q})$ in time and $O(2^{n_q})$ in memory, and that, rather than anything about the controller, is what bounds the size of this study. We would rather state the claim narrowly and have it hold: the mitigation layer is constant-cost in the register size, which is not the same as saying the simulation is.

\section{Experimental Setup}\label{exp}
Everything reported in this paper was produced by the released implementation, and this section records the conditions precisely enough to reproduce it. The software stack is Python 3.10.11 with Qiskit 1.4.0 and Qiskit Aer 0.16.1 for circuit execution, NumPy 2.2.6 and SciPy 1.14.1 for numerics, and scikit-learn 1.7.1 for the support-vector machine used by the kernel adapter. Mitigation techniques and the critical-difference calculation are implemented natively, so no external mitigation or post-hoc statistics package is required.

\begin{figure*}[h]
\centering
\includegraphics[width=\linewidth]{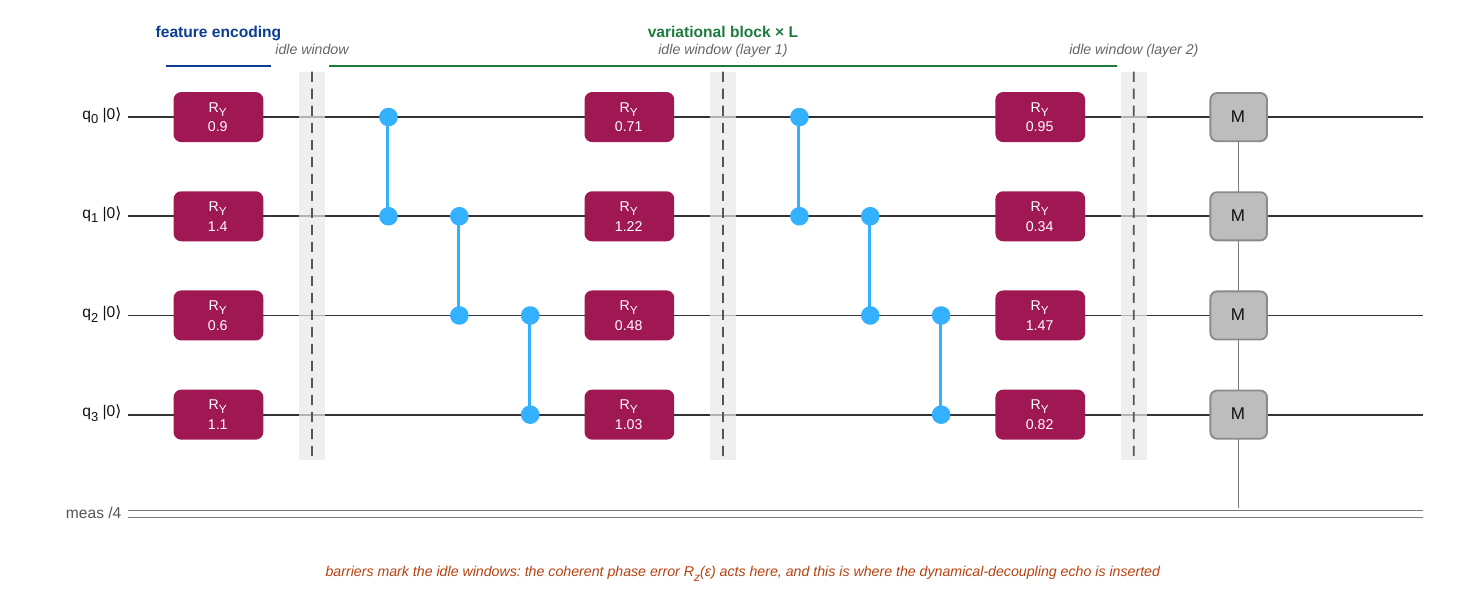}
\caption{The parameterized circuit shared by all five paradigms, rendered directly from the released implementation for $n_q=4$ and $L=2$. Each qubit is encoded by a single $R_y(x_i)$ rotation; every subsequent layer applies a nearest-neighbour chain of controlled-$Z$ gates followed by trainable $R_y(\theta_i)$ rotations. The barriers delimit the idle windows, which are the points at which the coherent phase error $R_z(\varepsilon)$ acts and, equivalently, the only places a dynamical-decoupling echo has anything to refocus. The circuit contains no mitigation-specific gates: APGEM selects which technique is applied to this circuit rather than modifying it, and the resulting circuit-level transformations are shown in Figure~\ref{fig:pool}.}
\label{fig:circuit}
\end{figure*}

The workload is the five adapters of Section~\ref{methodology}: a variational classifier, a quantum kernel machine, a variational eigensolver, a Born-machine generative model, and a quantum reinforcement-learning policy. All of them are built on the same four-qubit, two-layer ansatz of Figure~\ref{fig:circuit}, and each is evaluated against every technique in the pool, so a single number in the results tables is backed by the same circuit family across all five learners. Holding the circuit fixed is deliberate: it means any difference we report between paradigms reflects the learning task and the noise rather than an incidental choice of ansatz.

\begin{figure}[h]
\centering
\includegraphics[width=0.9\linewidth]{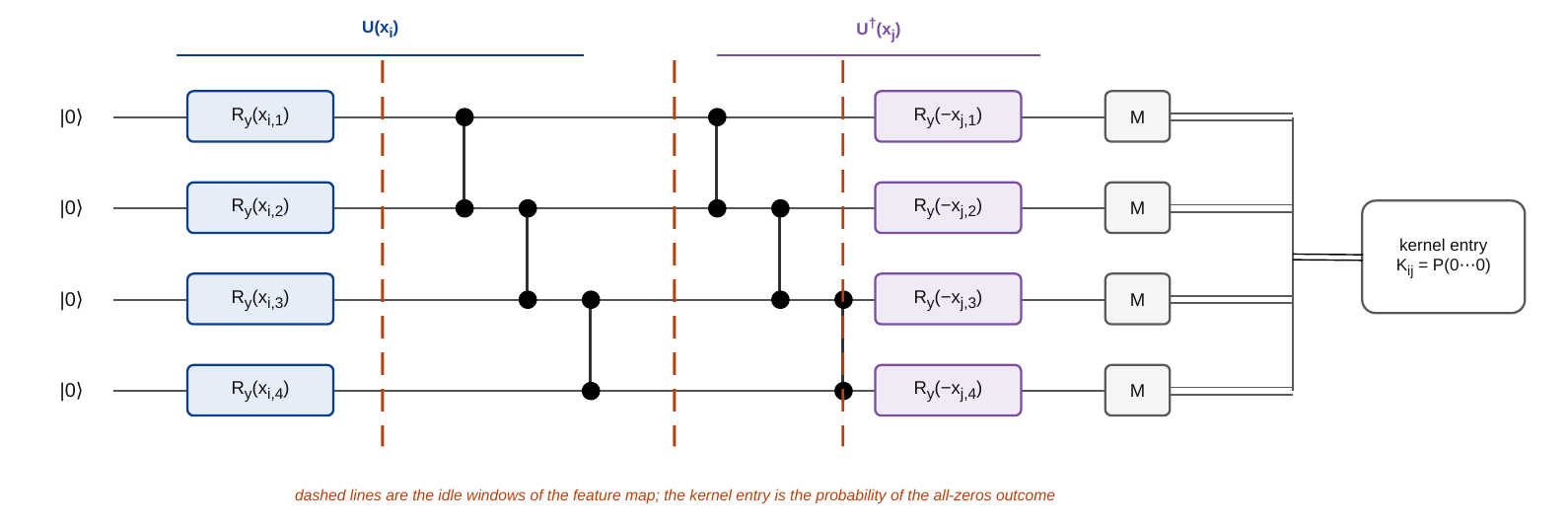}
\caption{Kernel circuit used by the QSVM adapter. The feature map $U(x_i)$ encodes a sample through $R_y$ rotations followed by a nearest-neighbour chain of controlled-$Z$ gates, and the inverse map $U^{\dagger}(x_j)$ is then applied for a second sample. The kernel entry is the probability of the all-zeros outcome, $K_{ij}=|\langle\phi(x_j)|\phi(x_i)\rangle|^{2}=P(0\cdots0)$, so noise in the measured distribution propagates directly into the kernel and from there into the downstream classifier. Dashed lines mark the idle windows of the feature map.}
\label{fig:kernel}
\end{figure}

 The kernel adapter is the one place where the circuit differs, since estimating a kernel entry requires composing a feature map with the inverse of another (Figure~\ref{fig:kernel}); the encoding and entangling structure is otherwise identical.

\begin{figure}[h!]
\centering
\includegraphics[width=\linewidth]{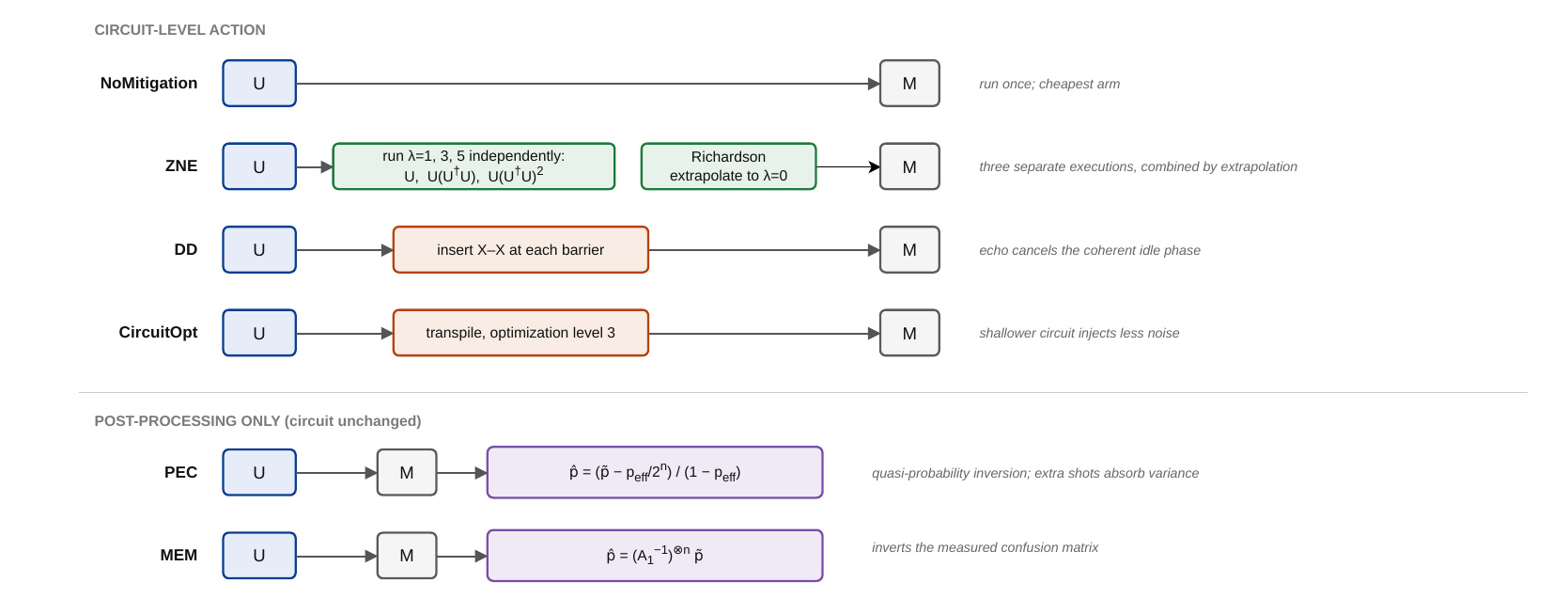}
\caption{What each arm of the mitigation pool does. Four act on the circuit before execution: ZNE folds the unitary at scale factors $\lambda\in\{1,3,5\}$ and extrapolates to the zero-noise limit, DD inserts an $X\!-\!X$ echo into every idle window so that a deterministic idle phase cancels, CircuitOpt recompiles to reduce the noise injected in the first place, and NoMitigation runs the circuit unchanged. The remaining two leave the circuit alone and correct the measured distribution: PEC inverts an effective depolarizing channel and MEM inverts the measured confusion matrix. Exactly one arm is selected per decision.}
\label{fig:pool}
\end{figure}

Six options are available at every decision: unmitigated execution, zero-noise extrapolation, probabilistic error cancellation, measurement error mitigation, dynamical decoupling, and compiler-level circuit optimization. Figure~\ref{fig:pool} shows what each of them does, which is worth being explicit about because the six split into two quite different kinds. Four transform the circuit before it runs, and two leave the circuit untouched and correct the measured distribution afterwards. The controller picks exactly one per decision. It never composes two techniques, and it never inserts corrective gates of its own, so any improvement it achieves comes from choosing well among existing methods rather than from adding a new correction mechanism.

Two families of operating conditions are used. The constructed grid crosses four depolarizing rates $\{0.01,0.05,0.12,0.18\}$ with two readout levels $\{0,0.10\}$ and two coherent idle rotations $\{0,0.5\}$, giving sixteen conditions per adapter and eighty task-condition blocks in total. Depolarizing and amplitude-damping errors attach to single-qubit gates, entangling gates carry twice the depolarizing rate, readout error is a symmetric bit-flip confusion matrix, and the coherent term is injected as a deterministic $R_z(\varepsilon)$ at each idle window rather than through the stochastic model. The four axes are varied independently rather than scaled together from a single parameter, because the whole question at issue is whether the controller can tell error \emph{classes} apart; tying readout error to depolarizing rate would make that impossible to test.

The second family replaces the constructed model entirely with calibration snapshots from eight IBM devices, listed in Table~\ref{tab:devices}. Each carries its own measured $T_1$ and $T_2$ times, per-gate error rates and per-qubit readout confusion matrices, and each is crossed with the coherent term to give sixteen further conditions. The spread across these devices is real but narrow compared to the constructed grid, and that turns out to matter for interpreting the results, so we report the ranges rather than a single representative device.

\begin{table}[htbp]
\centering
\caption{Range of measured parameters across the eight IBM calibration snapshots used for hardware-referenced validation.}
\label{tab:devices}
\begin{tabular}{lcccc}
\toprule
& Readout error & Two-qubit gate error & $T_1$ ($\mu$s) & $T_2$ ($\mu$s) \\
\midrule
Range across 8 devices & 0.014--0.037 & 0.006--0.013 & 62--148 & 39--115 \\
\bottomrule
\end{tabular}
\end{table}

Each evaluation draws 512 shots per circuit, and every task-condition-technique combination is executed repeatedly and cached in the response table described in Section~\ref{methodology}. Controllers are then replayed against that cached table, which is what makes the comparison paired: two controllers facing the same condition see the same distribution of realizations rather than two independent draws. The controller comparison uses a budget of 400 decisions over three seeds; the horizon study repeats it at 100, 400 and 1000 decisions to check convergence. The regret decomposition needs many repetitions at fixed context to separate expected utility from sampling noise, so it uses eight repetitions per technique per condition on the two adapters cheap enough to simulate at that depth, and four on the remaining three. Where a result rests on the smaller number we say so. The scalability sweep runs the eigensolver adapter at register sizes from three to eight qubits, measuring the cost of a complete six-technique decision at each size.

Alongside the adaptive controller we evaluate two alternative learners and six static policies behind the identical interface: Bayesian linear Thompson sampling, a context-free $\varepsilon$-greedy selector, and one fixed-technique controller for each arm of the pool. The static policies are not strawmen. They are exactly what current practice does, namely choose a technique once and apply it throughout, and the comparison is only meaningful because they face the same conditions from the same cached table.

Statistical testing treats the task-condition pair as the blocking factor and stays non-parametric. We run a Friedman omnibus test across the seven strategies, followed by a Nemenyi post-hoc test with its critical difference for the ranking, and pairwise Wilcoxon signed-rank tests of the controller against each fixed technique, each accompanied by Cliff's delta as an effect size. Means carry bootstrap $95\%$ confidence intervals over 5000 resamples. Parametric tests are avoided deliberately: utilities across conditions are neither independent nor normally distributed, and a $t$-test here would report a precision the data does not support.

\section{Results}
\label{sec:results}
We evaluate the controller on a factorial noise grid of $80$ device-condition blocks that spans five learning paradigms (VQC, QSVM, VQE, QGen, QRL) and six mitigation arms, and we additionally validate it on eight real IBM calibration snapshots that were held out from training. Throughout, utility is the cost-aware sigmoid of output quality defined in Section~\ref{exp}. We define the \emph{oracle} as the per-block expected-utility argmax over the six arms, and \emph{oracle match} as the fraction of blocks in which the controller picks that same arm. Unless noted otherwise, confidence intervals are $95\%$ bootstrap intervals computed over blocks, and every run shares the fixed software stack listed in Table~\ref{tab:env}.

\begin{table}[htbp]
\centering
\caption{Software environment. All experiments were executed under a single fixed stack to keep the noise models and the numerical pipeline reproducible.}
\label{tab:env}
\begin{tabular}{ll}
\toprule
Component & Version \\
\midrule
Python        & 3.10.11 \\
Qiskit        & 1.4.0 \\
Qiskit Aer    & 0.16.1 \\
NumPy         & 2.2.6 \\
SciPy         & 1.14.1 \\
scikit-learn  & 1.7.1 \\
Matplotlib    & 3.10.0 \\
\bottomrule
\end{tabular}
\end{table}

\subsection{The controller outperforms every static policy}
\label{sec:r-main}
\begin{figure}[htbp]
\centering
\includegraphics[width=0.7\linewidth]{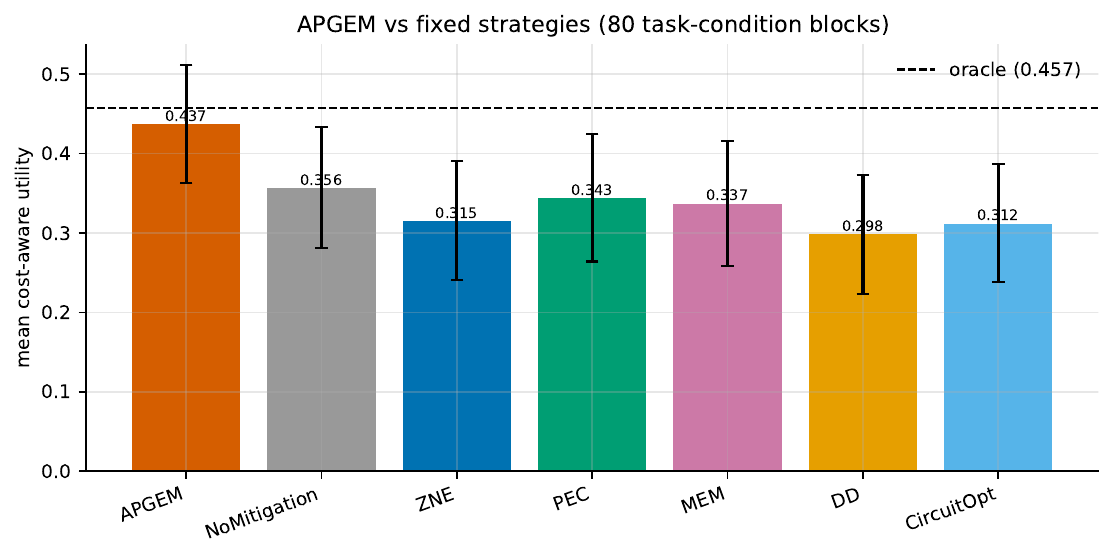}
\caption{Mean utility of the controller against the six static mitigation policies over the $80$-block grid, with $95\%$ bootstrap intervals. The controller exceeds every static arm and tracks the oracle.}
\label{fig:vsstatic}
\end{figure}

Over the full $80$-block grid, the contextual controller reaches a mean utility of $0.435$. That beats $0.356$ for the best static policy (NoMitigation) and $0.298$ for the worst (DD), and it recovers $95.5\%$ of the $0.457$ mean oracle utility (Fig.~\ref{fig:vsstatic}). None of the fixed policies come close: even NoMitigation, the strongest of them on average, only matches the oracle's choice on $40\%$ of blocks. The controller also cuts expected regret substantially, down to $0.0755$ versus $0.1037$ for NoMitigation and $0.1629$ for DD (Table~\ref{tab:main}). The fact that the static arms rank so differently from one another is itself telling: no single fixed strategy wins across the board, which is precisely the gap the controller is designed to close.

\begin{table}[htbp]
\centering
\caption{Overall comparison over the $80$-block grid. The two context-aware controllers lead every static policy and the context-free selector on mean utility and regret. Rows are ordered by mean utility.}
\label{tab:main}
\begin{tabular}{lcccc}
\toprule
Method & Mean utility & $95\%$ CI & Oracle match & Mean regret \\
\midrule
\textbf{APGEM (LinUCB)} & \textbf{0.435} & [0.323, 0.542] & \textbf{0.558} & 0.0755 \\
Thompson sampling       & 0.431 & [0.315, 0.541] & 0.533 & \textbf{0.0587} \\
$\varepsilon$-greedy    & 0.398 & [0.280, 0.508] & 0.508 & 0.0761 \\
Static NoMitigation     & 0.356 & [0.237, 0.477] & 0.400 & 0.1037 \\
Static PEC              & 0.343 & [0.200, 0.468] & 0.263 & 0.1125 \\
Static MEM              & 0.337 & [0.221, 0.454] & 0.087 & 0.1227 \\
Static ZNE              & 0.315 & [0.200, 0.426] & 0.100 & 0.1417 \\
Static CircuitOpt       & 0.312 & [0.200, 0.427] & 0.050 & 0.1473 \\
Static DD               & 0.298 & [0.188, 0.415] & 0.100 & 0.1629 \\
\bottomrule
\end{tabular}
\end{table}

\begin{figure}[htbp]
\centering
\includegraphics[width=0.85\linewidth]{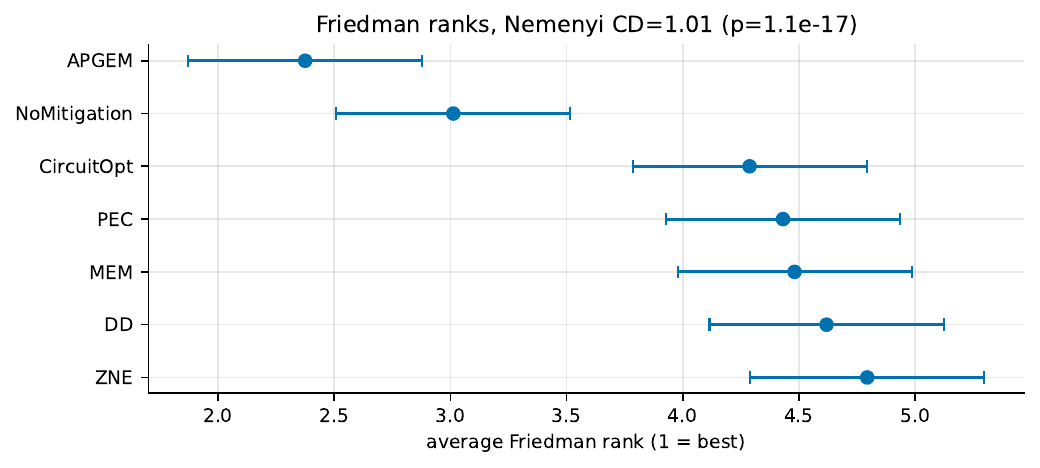}
\caption{Critical-difference diagram (Nemenyi, $\text{CD}=1.007$) over the seven strategies. APGEM holds the best average rank and is separated from every static arm.}
\label{fig:cd}
\end{figure}

This advantage holds up statistically. A Friedman test over the seven strategies (APGEM plus the six static arms) rejects equal ranks ($\chi^{2}=92.18$, $p=1.07\times10^{-17}$), and APGEM comes out with the best mean rank, $2.375$ (Table~\ref{tab:sig}). Post hoc Nemenyi comparisons at $\text{CD}=1.007$ separate APGEM from every static arm, shown as a critical-difference diagram in Fig.~\ref{fig:cd}. Pairwise Wilcoxon signed-rank tests against APGEM are significant across all six comparisons; the effect is smallest against NoMitigation (Cliff's $\delta=0.101$, $p=3.8\times10^{-6}$) and largest against ZNE (Cliff's $\delta=0.232$, $p=8.2\times10^{-12}$).

\begin{table}[htbp]
\centering
\caption{Significance of APGEM over each static arm on the $80$-block grid. Average ranks are from the Friedman test ($\chi^{2}=92.18$, $p=1.07\times10^{-17}$, Nemenyi $\text{CD}=1.007$). $p$-values are Wilcoxon signed-rank against APGEM; $\delta$ is Cliff's delta.}
\label{tab:sig}
\begin{tabular}{lccc}
\toprule
Method & Avg. rank & Wilcoxon $p$ & Cliff's $\delta$ \\
\midrule
\textbf{APGEM} & \textbf{2.375} & --- & --- \\
NoMitigation   & 3.013 & $3.8\times10^{-6}$  & 0.101 \\
CircuitOpt     & 4.287 & $7.0\times10^{-9}$  & 0.176 \\
PEC            & 4.431 & $8.4\times10^{-9}$  & 0.156 \\
MEM            & 4.481 & $2.1\times10^{-8}$  & 0.172 \\
DD             & 4.619 & $5.6\times10^{-9}$  & 0.192 \\
ZNE            & 4.794 & $8.2\times10^{-12}$ & 0.232 \\
\bottomrule
\end{tabular}
\end{table}

\subsection{One controller learns five distinct policies}
\label{sec:r-para}
\begin{figure}[htbp]
\centering
\includegraphics[width=0.8\linewidth]{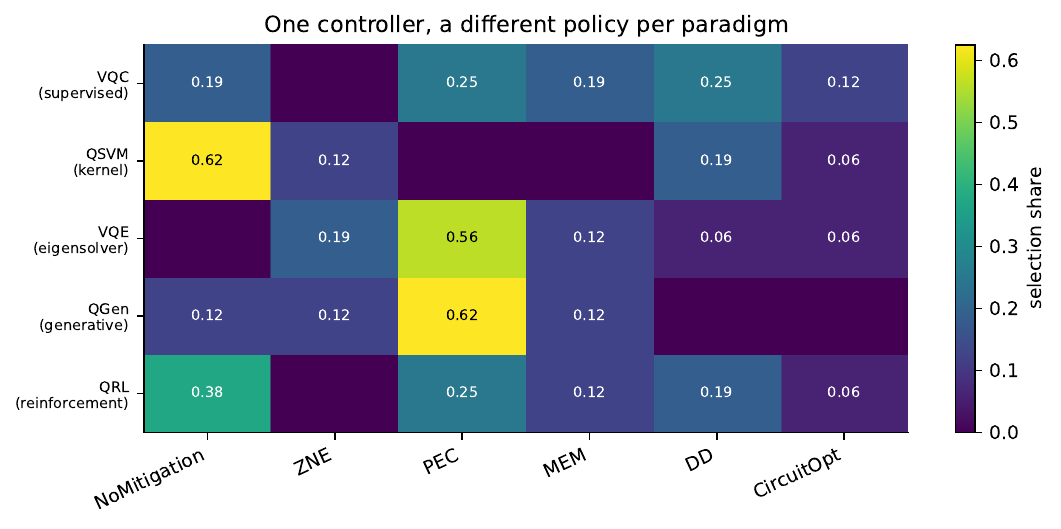}
\caption{Learned selection share per paradigm. One controller, five distinct policies, inferred from the shared feedback interface without knowledge of the learning task.}
\label{fig:paradigm}
\end{figure}

What is striking is that a single controller, working through one shared feedback interface and blind to which task it is facing, still converges to a different selection policy for each paradigm (Table~\ref{tab:para}, Fig.~\ref{fig:paradigm}). On the deterministic eigensolver (VQE) it settles on PEC ($0.56$ share of blocks) and reaches $98.9\%$ of the paradigm's oracle utility; on the generative task (QGen) it again leans on PEC ($0.62$) and reaches $99.3\%$. On the kernel classifier (QSVM), where mitigation typically is not worth its cost, the controller correctly defaults to NoMitigation ($0.62$). The variational classifier (VQC) is the one case with a genuinely mixed optimum, and the controller's choices reflect that: it splits its mass fairly evenly between DD and PEC ($0.25$ each). Argmax match is lowest for QRL ($0.25$), yet the controller still captures $94\%$ of that paradigm's oracle utility. As we show in Section~\ref{sec:r-regret}, these apparent disagreements are mostly near-ties produced by QRL's stochastic routing, not genuine selection errors.

\begin{table}[htbp]
\centering
\caption{Per-paradigm outcome. The controller learns a distinct dominant choice per paradigm and recovers a high fraction of the per-paradigm oracle utility. Argmax match is lowest for QRL, whose stochastic policy makes many blocks near-ties (Section~\ref{sec:r-regret}).}
\label{tab:para}
\begin{tabular}{llccc}
\toprule
Paradigm & Dominant choice (share) & APGEM util. & Oracle util. & Argmax match \\
\midrule
VQC  & DD / PEC (0.25 each)  & 0.465 & 0.497 & 0.38 \\
QSVM & NoMitigation (0.62)   & 0.078 & 0.095 & 0.69 \\
VQE  & PEC (0.56)            & 0.370 & 0.375 & 0.75 \\
QGen & PEC (0.62)            & 0.526 & 0.530 & 0.81 \\
QRL  & NoMitigation (0.38)   & 0.743 & 0.790 & 0.25 \\
\bottomrule
\end{tabular}
\end{table}

\subsection{Selection tracks the noise regime}
\label{sec:r-regime}
The controller's policy tracks more than just the paradigm; it also reads the noise conditions. Table~\ref{tab:regime} bins the blocks into six noise regimes and compares the controller's dominant arm against the oracle's in each. At very low noise it leans on NoMitigation, at low noise it shifts to ZNE, and in the moderate regime it settles on PEC ($0.60$ share, close to the oracle's $0.55$). When noise turns severe or readout-dominated, it falls back to NoMitigation, matching the oracle, since the heavier adapters no longer earn back their sampling cost. The one place it visibly departs from the oracle is the coherent-dominated regime, where the controller favours PEC while the oracle marginally prefers NoMitigation. The two options are close enough in expected utility there that this disagreement barely shows up as regret.

\begin{table}[htbp]
\centering
\caption{Dominant selection by noise regime. The controller's top choice tracks the oracle's top choice in every regime except the coherent-dominated one, where the two arms are near-ties.}
\label{tab:regime}
\begin{tabular}{lccc}
\toprule
Noise regime & Blocks & APGEM top (share) & Oracle top (share) \\
\midrule
Very low            & 5  & NoMitigation (0.40) & NoMitigation (1.00) \\
Low                 & 5  & ZNE (0.40)          & ZNE (0.60) \\
Moderate            & 20 & PEC (0.60)          & PEC (0.55) \\
Severe              & 20 & NoMitigation (0.40) & NoMitigation (0.55) \\
Readout-dominated   & 10 & NoMitigation (0.30) & NoMitigation (0.40) \\
Coherent-dominated  & 20 & PEC (0.30)          & NoMitigation (0.35) \\
\bottomrule
\end{tabular}
\end{table}

\begin{figure}[htbp]
\centering
\includegraphics[width=\linewidth]{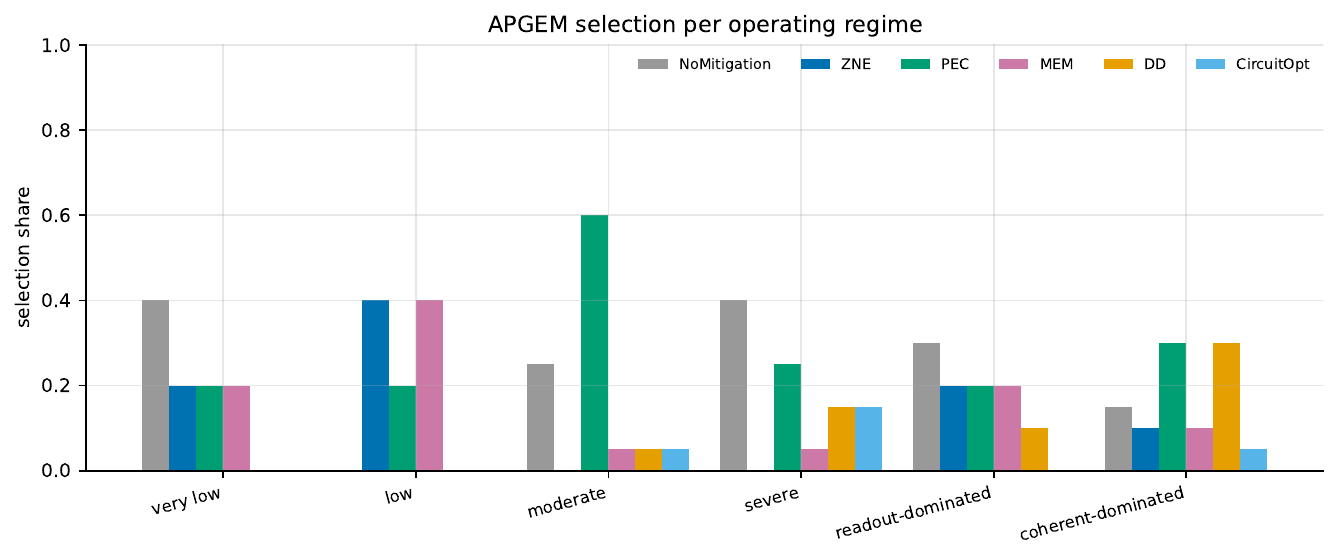}
\caption{Selection share by noise regime for the controller (left) and the oracle (right). The controller reproduces the oracle's regime structure.}
\label{fig:regime}
\end{figure}

\begin{figure}[htbp]
\centering
\includegraphics[width=\linewidth]{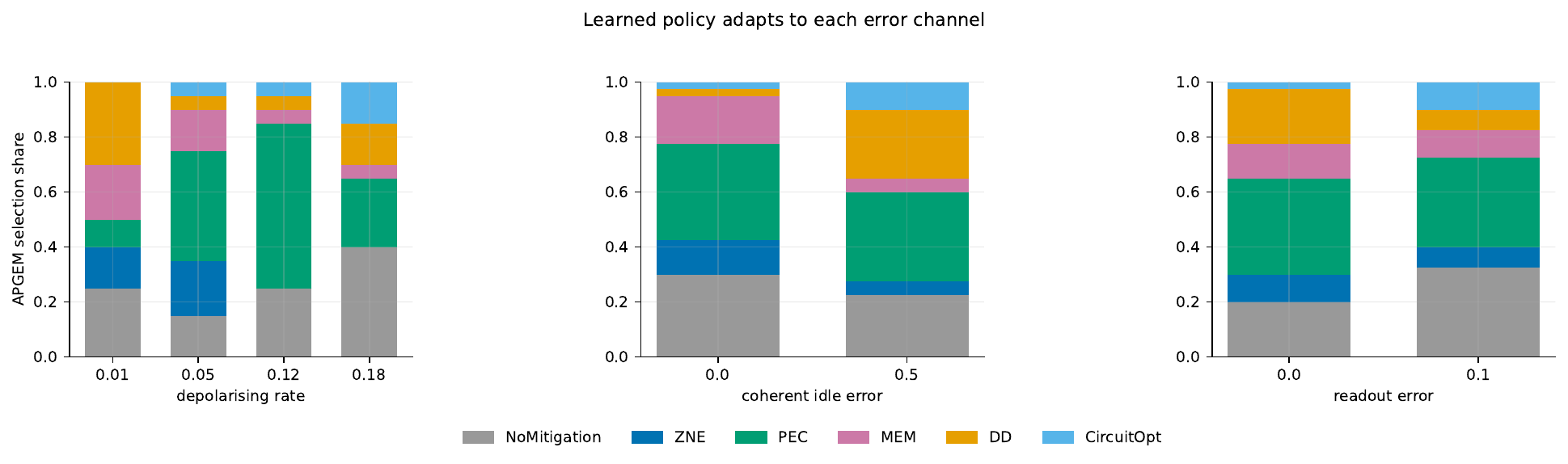}
\caption{Selection share as a function of the depolarizing, readout, and coherent noise axes. PEC dominates the mid-noise band and yields to NoMitigation at the extremes.}
\label{fig:selnoise}
\end{figure}

\begin{figure}[htbp]
\centering
\includegraphics[width=0.85\linewidth]{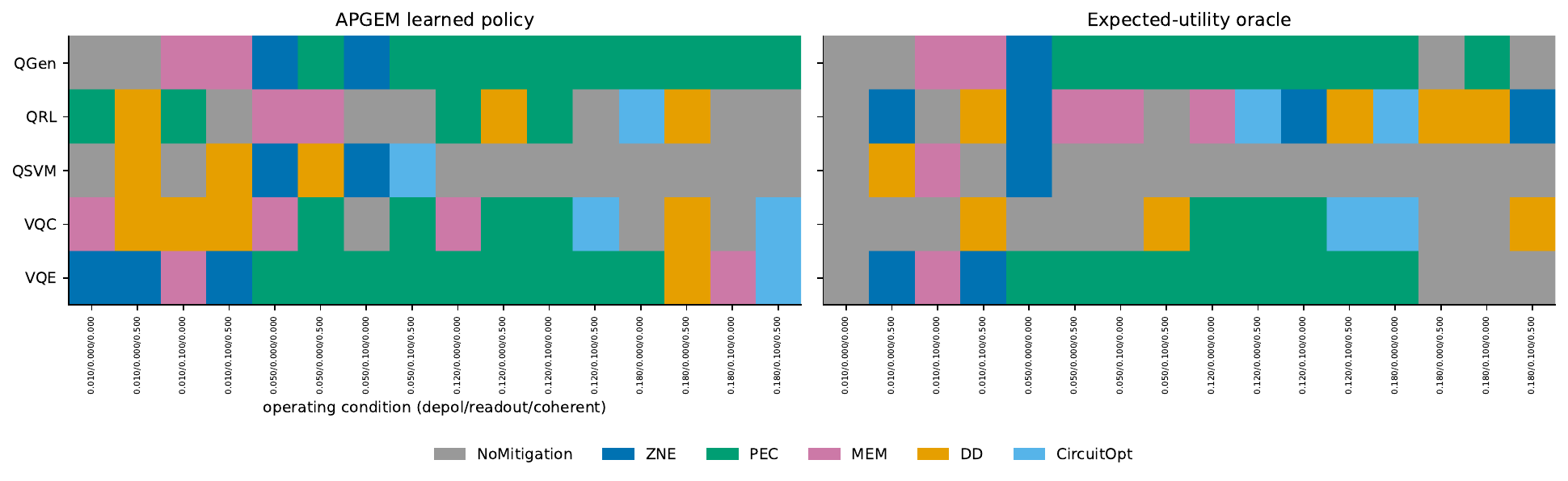}
\caption{Learned policy map over the depolarizing and readout error axes. Each cell shows the controller's dominant arm, revealing contiguous decision regions rather than a single global choice.}
\label{fig:policymap}
\end{figure}

\begin{figure}[htbp]
\centering
\includegraphics[width=\linewidth]{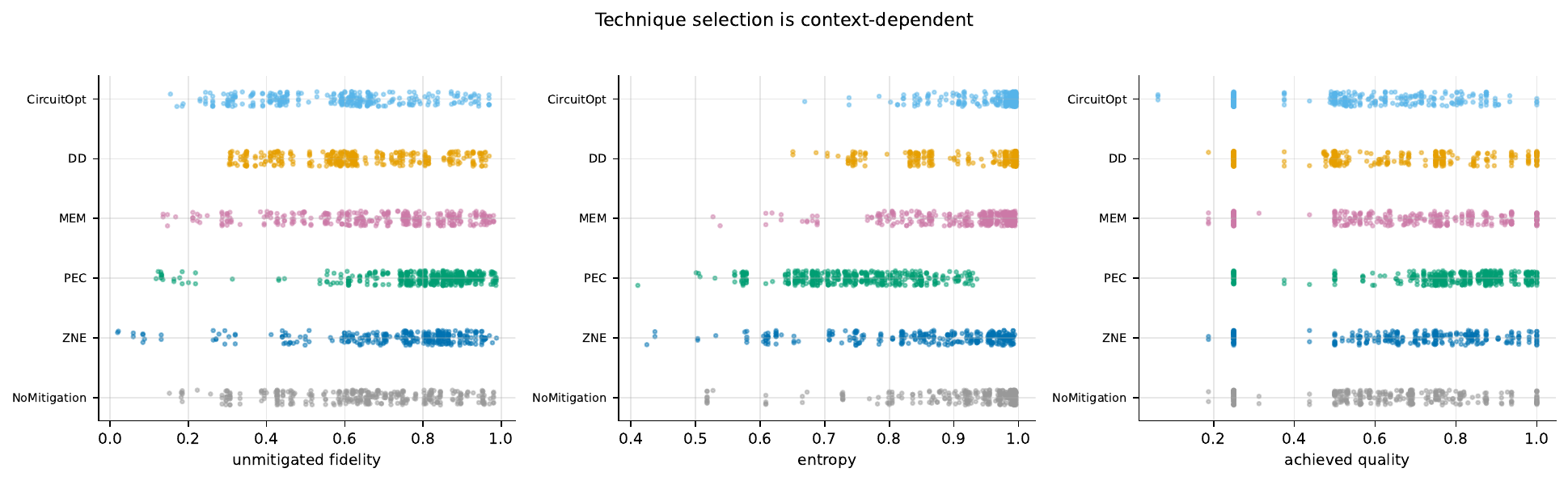}
\caption{Selection projected onto the raw context features observed by the controller. The decision boundaries align with the physically meaningful features (two-qubit error, readout error, circuit depth).}
\label{fig:selcontext}
\end{figure}

\begin{figure}[htbp]
\centering
\includegraphics[width=0.7\linewidth]{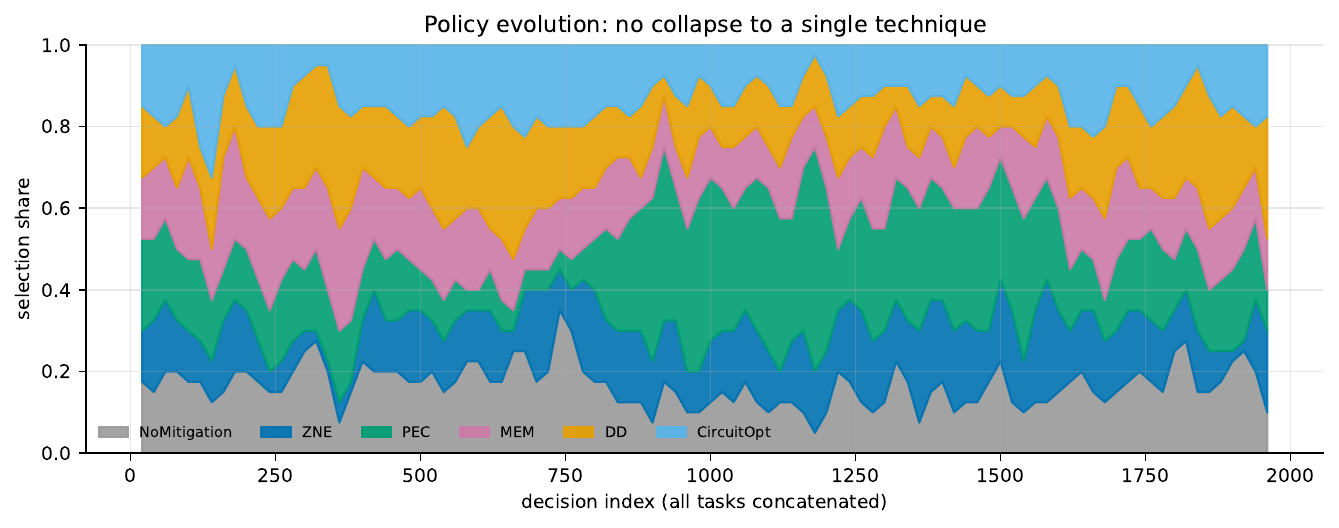}
\caption{Evolution of the selection distribution over interactions. The mix stabilises as the controller's per-arm models are refined.}
\label{fig:seltime}
\end{figure}

Fig.~\ref{fig:regime} plots this regime dependence directly, contrasting the controller's and the oracle's selection shares side by side, while Fig.~\ref{fig:selnoise} breaks it down along the individual noise axes: PEC usage climbs with the depolarizing rate up through the moderate band, then falls away as NoMitigation takes back over at the extreme -- exactly the cost-aware behaviour we would want. Fig.~\ref{fig:policymap} lays out the full learned policy as a map over the depolarizing and readout axes, and Fig.~\ref{fig:selcontext} projects the same selections onto the raw context features the controller actually sees. Finally, Fig.~\ref{fig:seltime} tracks how these choices evolve over time: the selection distribution settles down as interactions accumulate, which points to genuine convergence rather than ongoing churn.

\subsection{Most of the residual gap is sampling noise, not policy error}
\label{sec:r-regret}
\begin{figure}[htbp]
\centering
\includegraphics[width=\linewidth]{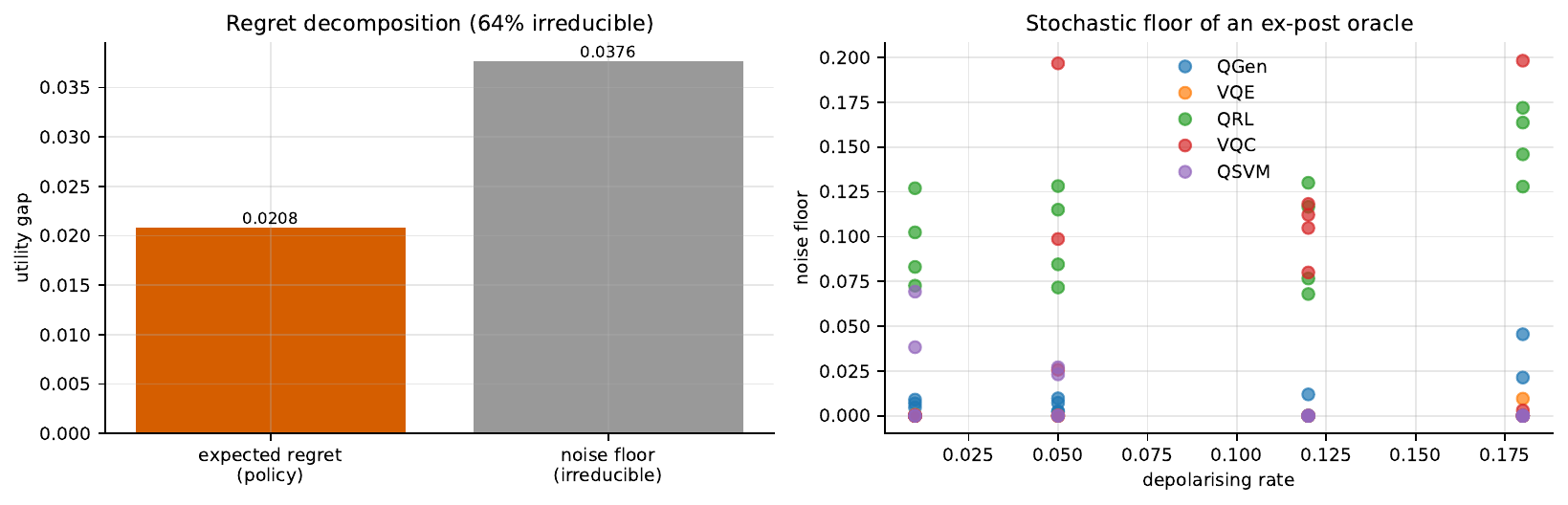}
\caption{Pooled regret decomposition. The controller's expected regret against the expected-utility oracle sits well below the irreducible sampling-noise floor of the ex-post oracle.}
\label{fig:regret}
\end{figure}

\begin{figure}[htbp]
\centering
\includegraphics[width=\linewidth]{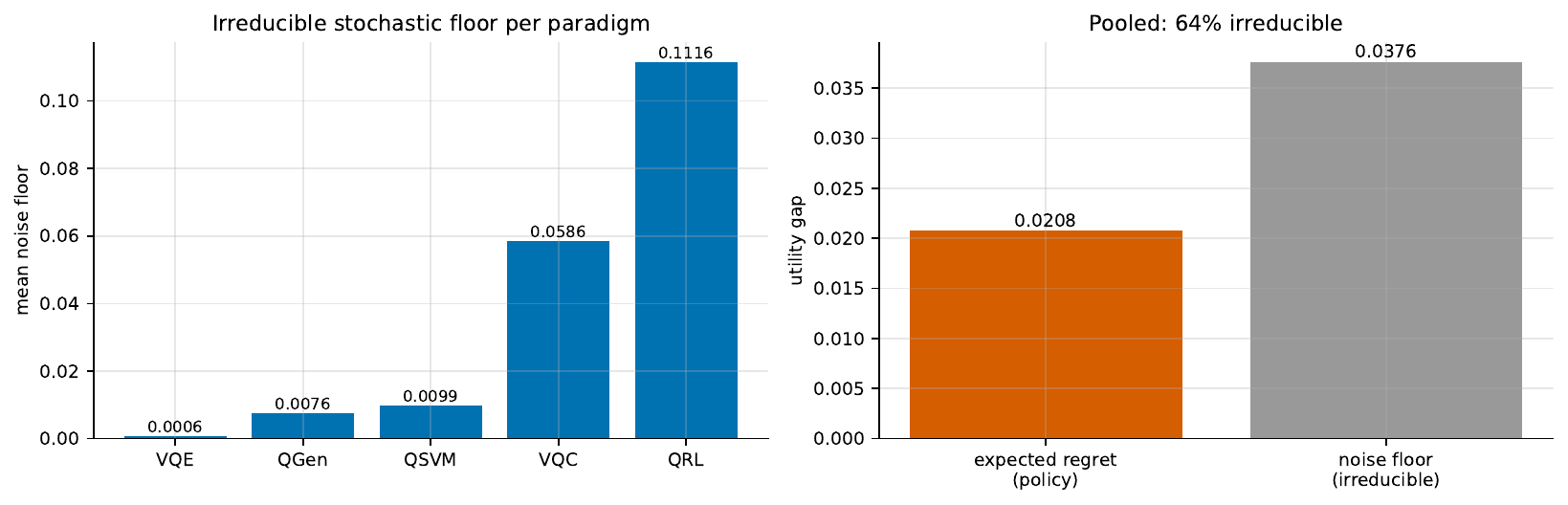}
\caption{Regret decomposition per paradigm. Left: irreducible noise floor per paradigm. Right: pooled expected regret against the irreducible floor.}
\label{fig:regretall}
\end{figure}

There is a gap between the controller capturing $95.5\%$ of oracle utility and only matching the exact argmax on $57.5\%$ of blocks, and this subsection is about explaining where that gap comes from. Fig.~\ref{fig:regret} breaks the regret down. Against the expected-utility oracle, mean expected regret is $0.0208$; but the irreducible floor set by a per-realization ex-post oracle is $0.0376$. In other words, roughly $64\%$ of what a naive after-the-fact comparison would blame on the controller is really just sampling noise that no policy could have avoided. Table~\ref{tab:regret} breaks this floor down by paradigm, and it spans two orders of magnitude -- from $0.0006$ for the deterministic eigensolver up to $0.1116$ for the stochastic reinforcement-learning task, whose sampled routing decisions make any single realization noisy. That is also the reason QRL has the lowest argmax match in Table~\ref{tab:para} even though it still reaches $94\%$ of its oracle utility: the disagreements there are near-ties, not real mistakes. Fig.~\ref{fig:regretall} shows the same breakdown for all five paradigms.

\begin{table}[htbp]
\centering
\caption{Regret decomposition. The irreducible per-realization sampling-noise floor varies by two orders of magnitude across paradigms and dominates the controller's own expected regret. Pooled expected regret is $0.0208$ against a pooled floor of $0.0376$.}
\label{tab:regret}
\begin{tabular}{lccccc}
\toprule
 & VQE & QGen & QSVM & VQC & QRL \\
\midrule
Noise floor & 0.0006 & 0.0076 & 0.0099 & 0.0586 & 0.1116 \\
\bottomrule
\end{tabular}
\end{table}

\subsection{Context, not the bandit, drives the gain}
\label{sec:r-ctrl}
To check that the gain comes from conditioning on context rather than from our choice of bandit algorithm, we ran the same comparison with a Bayesian linear Thompson-sampling controller and a context-free $\varepsilon$-greedy selector alongside the six static policies, all behind the same interface (Table~\ref{tab:main}). The two contextual controllers land close together -- $0.435$ for LinUCB versus $0.431$ for Thompson sampling -- and both clearly beat the context-free selector ($0.398$) as well as every static policy. That pattern suggests the gain comes from the framework and its context representation rather than from any particular bandit; either contextual algorithm would do the job. Thompson sampling does edge out LinUCB on mean regret ($0.0587$ versus $0.0755$), so a deployment that cares more about regret than raw utility might prefer it, but this does not change the overall conclusion.

\subsection{The mitigation arms are genuinely different}
\label{sec:r-tech}
\begin{figure}[htbp]
\centering
\includegraphics[width=0.7\linewidth]{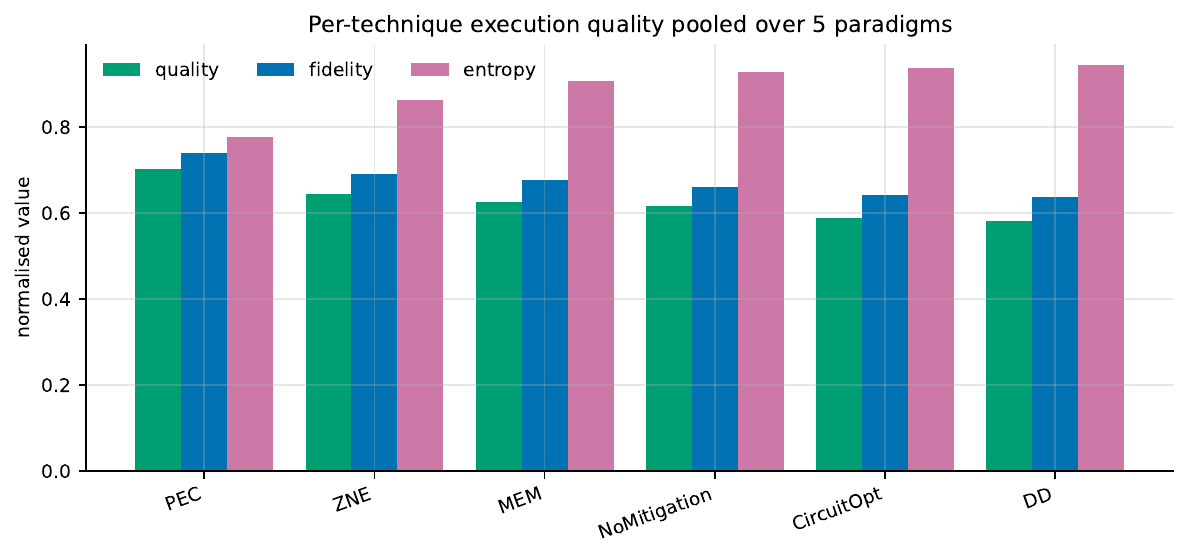}
\caption{Pooled output quality per mitigation technique with $95\%$ bootstrap intervals. The arms occupy distinct quality bands.}
\label{fig:techquality}
\end{figure}

\begin{figure}[htbp]
\centering
\includegraphics[width=0.85\linewidth]{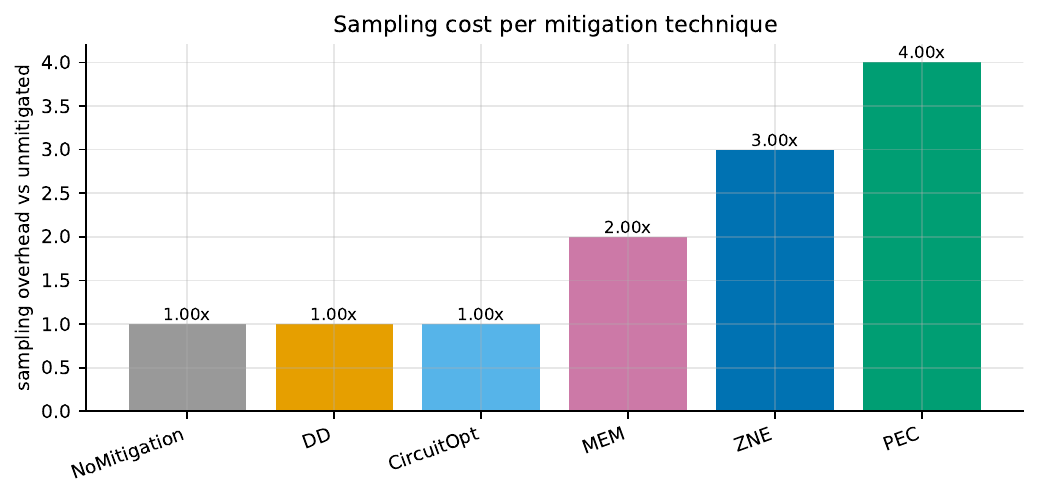}
\caption{Sampling overhead per technique relative to the NoMitigation budget. PEC and ZNE buy their quality with a $4\times$ and $3\times$ shot cost respectively.}
\label{fig:overhead}
\end{figure}

For the controller's job to mean anything, the arms it is choosing between need to actually behave differently -- and they do. Table~\ref{tab:tech} characterizes each mitigation technique pooled across all blocks: output quality, state fidelity, output entropy, per-decision shot budget, sampling overhead, and projected hardware time. PEC gives the best pooled quality ($0.702$) and fidelity ($0.741$) and the lowest entropy ($0.777$), but it costs $4\times$ the shots; DD and CircuitOpt stay at the base $1\times$ budget but come with the lowest quality of the six. Fig.~\ref{fig:techquality} and Fig.~\ref{fig:overhead} plot this quality-versus-cost spread, and it is exactly the trade-off the cost-aware utility function is built to capture -- which is also why no single arm wins in every situation.

\begin{table}[htbp]
\centering
\caption{Technique characterization pooled over all blocks. Quality is reported with its $95\%$ bootstrap interval. Overhead is relative to the NoMitigation shot budget; projected hardware time assumes a fixed per-shot rate.}
\label{tab:tech}
\begin{tabular}{lcccccc}
\toprule
Technique & Quality [95\% CI] & Fidelity & Entropy & Shots & Overhead & HW (s) \\
\midrule
PEC          & 0.702 [0.666, 0.738] & 0.741 & 0.777 & 26214 & $4.0\times$ & 10.49 \\
ZNE          & 0.645 [0.611, 0.681] & 0.692 & 0.864 & 19661 & $3.0\times$ & 7.86 \\
MEM          & 0.627 [0.593, 0.661] & 0.678 & 0.907 & 13107 & $2.0\times$ & 5.24 \\
NoMitigation & 0.618 [0.585, 0.651] & 0.661 & 0.928 & 6554  & $1.0\times$ & 2.62 \\
CircuitOpt   & 0.589 [0.556, 0.622] & 0.642 & 0.939 & 6554  & $1.0\times$ & 2.62 \\
DD           & 0.583 [0.550, 0.616] & 0.637 & 0.944 & 6554  & $1.0\times$ & 2.62 \\
\bottomrule
\end{tabular}
\end{table}

\subsection{Robustness to increasing noise}
\label{sec:r-noise}
\begin{figure}[htbp]
\centering
\includegraphics[width=0.7\linewidth]{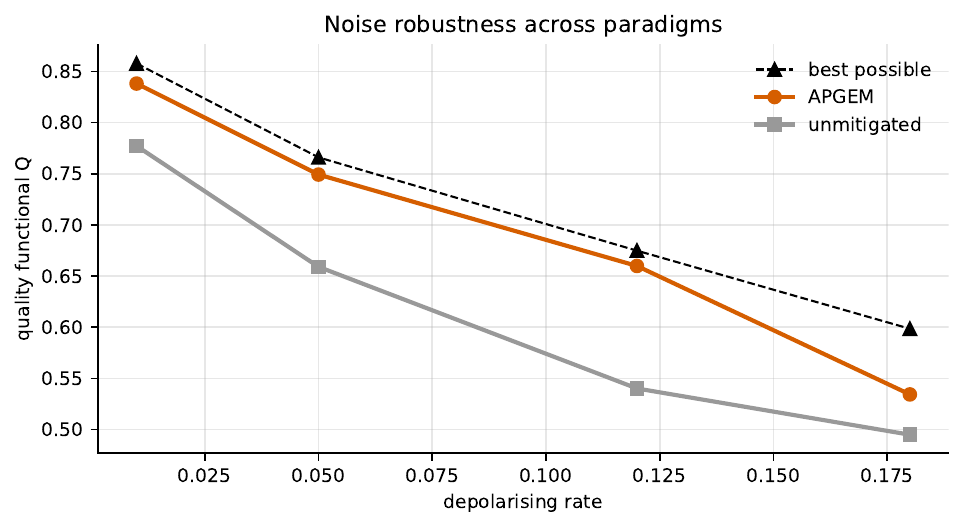}
\caption{Controller utility versus depolarizing rate, against the NoMitigation baseline and the best-possible arm. The controller tracks the achievable frontier.}
\label{fig:noiserobust}
\end{figure}

Table~\ref{tab:noise} and Fig.~\ref{fig:noiserobust} sweep the depolarizing rate from $0.01$ up to $0.18$. Across this whole range the controller stays close to the best achievable utility and keeps a consistent edge over the NoMitigation baseline. At the mild end ($0.01$) it reaches $0.838$ against a best-possible $0.858$; at the severe end ($0.18$) it reaches $0.534$ against $0.598$, while NoMitigation drops off to $0.495$. Its margin over the baseline is widest in the mid-noise band -- which happens to be where picking the right mitigation matters most.

\begin{table}[htbp]
\centering
\caption{Utility versus depolarizing rate. The controller tracks the best-possible arm and holds a margin over the NoMitigation baseline across the range.}
\label{tab:noise}
\begin{tabular}{lccc}
\toprule
Depol. rate & APGEM & NoMitigation & Best possible \\
\midrule
0.01 & 0.838 & 0.777 & 0.858 \\
0.05 & 0.749 & 0.659 & 0.766 \\
0.12 & 0.660 & 0.540 & 0.675 \\
0.18 & 0.534 & 0.495 & 0.598 \\
\bottomrule
\end{tabular}
\end{table}

\subsection{Scalability}
\label{sec:r-scal}
\begin{figure}[htbp]
\centering
\includegraphics[width=\linewidth]{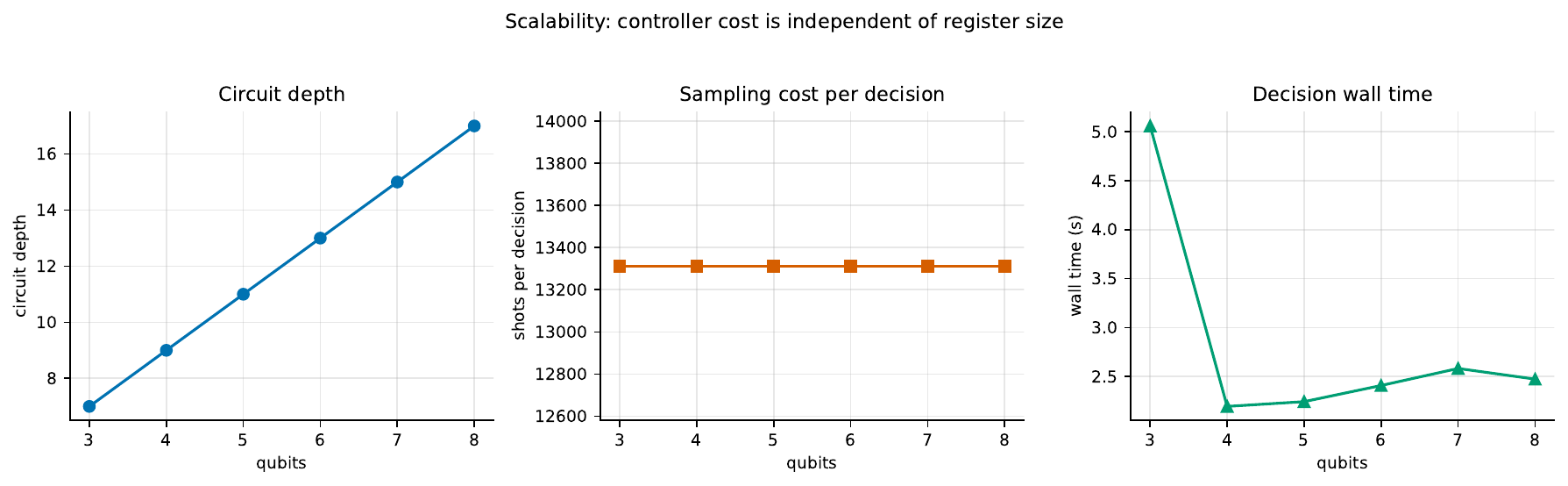}
\caption{Best-arm quality and per-decision wall time against qubit count. Decision cost is flat; quality degrades gracefully with circuit size.}
\label{fig:scal}
\end{figure}

Table~\ref{tab:scal} and Fig.~\ref{fig:scal} report a scaling sweep from three to eight qubits, with circuit depth and two-qubit gate count growing alongside qubit count. Because the per-decision shot budget is fixed at $13312$ shots, the controller's decision cost does not grow with circuit size. Wall time per decision holds steady, roughly two to five seconds across the whole range, and the achieved best-arm quality degrades gracefully, from $0.893$ at three qubits down to $0.810$ at eight. That decline tracks the growing circuit noise itself rather than any breakdown in how the controller selects.

\begin{table}[htbp]
\centering
\caption{Scaling sweep. The per-decision shot budget is fixed, so decision cost is constant in circuit size; achieved quality degrades gracefully as the circuit grows.}
\label{tab:scal}
\begin{tabular}{lccccccc}
\toprule
Qubits & Depth & 2Q gates & Params & Decision shots & Best arm & Quality & Wall (s) \\
\midrule
3 & 7  & 4  & 6  & 13312 & PEC & 0.893 & 5.06 \\
4 & 9  & 6  & 8  & 13312 & PEC & 0.872 & 2.19 \\
5 & 11 & 8  & 10 & 13312 & PEC & 0.885 & 2.24 \\
6 & 13 & 10 & 12 & 13312 & PEC & 0.875 & 2.41 \\
7 & 15 & 12 & 14 & 13312 & PEC & 0.862 & 2.58 \\
8 & 17 & 14 & 16 & 13312 & PEC & 0.810 & 2.47 \\
\bottomrule
\end{tabular}
\end{table}

\subsection{Learning dynamics and interaction horizon}
\label{sec:r-horizon}
\begin{figure}[htbp]
\centering
\includegraphics[width=\linewidth]{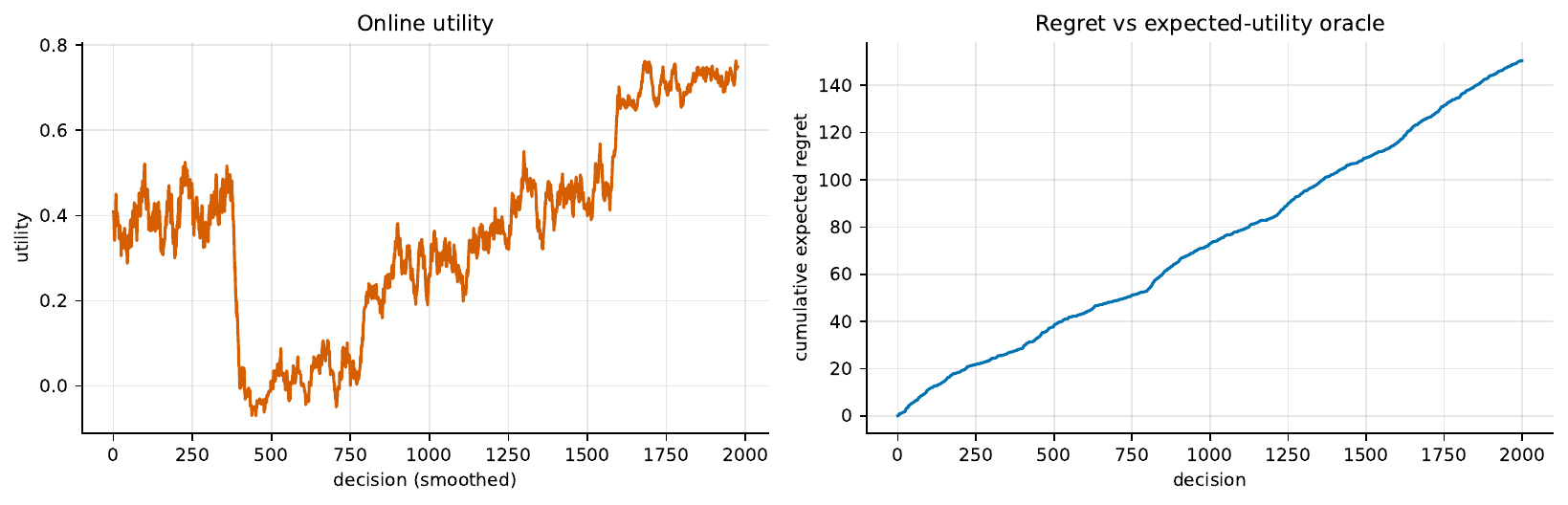}
\caption{Learning curves. Utility rises and expected regret decays with the number of interactions across the five paradigms.}
\label{fig:learning}
\end{figure}

\begin{figure}[htbp]
\centering
\includegraphics[width=\linewidth]{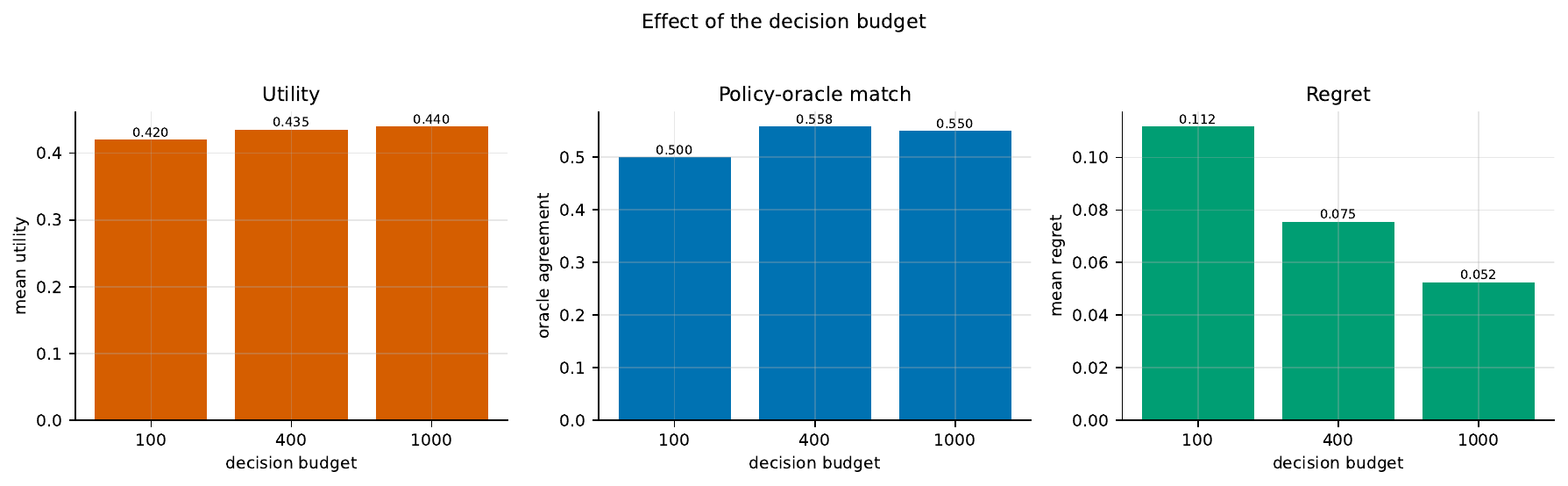}
\caption{Effect of the interaction horizon on utility, oracle match, and regret. Regret falls steadily while utility saturates by the $400$-interaction setting used for the main results.}
\label{fig:horizons}
\end{figure}

The controller gets better with more experience. Table~\ref{tab:horizon} varies the interaction horizon across $100$, $400$, and $1000$ interactions. Mean utility climbs from $0.420$ to $0.440$, but the more telling number is mean regret, which falls monotonically from $0.112$ to $0.052$ as the per-arm models sharpen -- at the cost of a proportional increase in total shot budget. Fig.~\ref{fig:learning} shows the corresponding learning curves: utility rises and regret decays with more interactions across all five paradigms, and the selection distribution settles down as the horizon lengthens (Fig.~\ref{fig:horizons}). We use $400$ interactions for the main results because it sits on the near-flat part of the utility curve while keeping the shot budget an order of magnitude below what the $1000$-interaction setting requires.

\begin{table}[htbp]
\centering
\caption{Sensitivity to interaction horizon. Regret falls monotonically with more interactions while utility saturates; total shot budget grows proportionally.}
\label{tab:horizon}
\begin{tabular}{lcccc}
\toprule
Horizon & Mean utility [95\% CI] & Oracle match & Mean regret & Total shots \\
\midrule
100  & 0.420 [0.309, 0.527] & 0.500 & 0.1117 & $1.28\times10^{6}$ \\
400  & 0.435 [0.323, 0.542] & 0.558 & 0.0755 & $4.85\times10^{6}$ \\
1000 & 0.440 [0.327, 0.546] & 0.550 & 0.0525 & $1.20\times10^{7}$ \\
\bottomrule
\end{tabular}
\end{table}

\subsection{Validation on real device calibration data}
\label{sec:r-cal}
\begin{figure}[htbp]
\centering
\includegraphics[width=\linewidth]{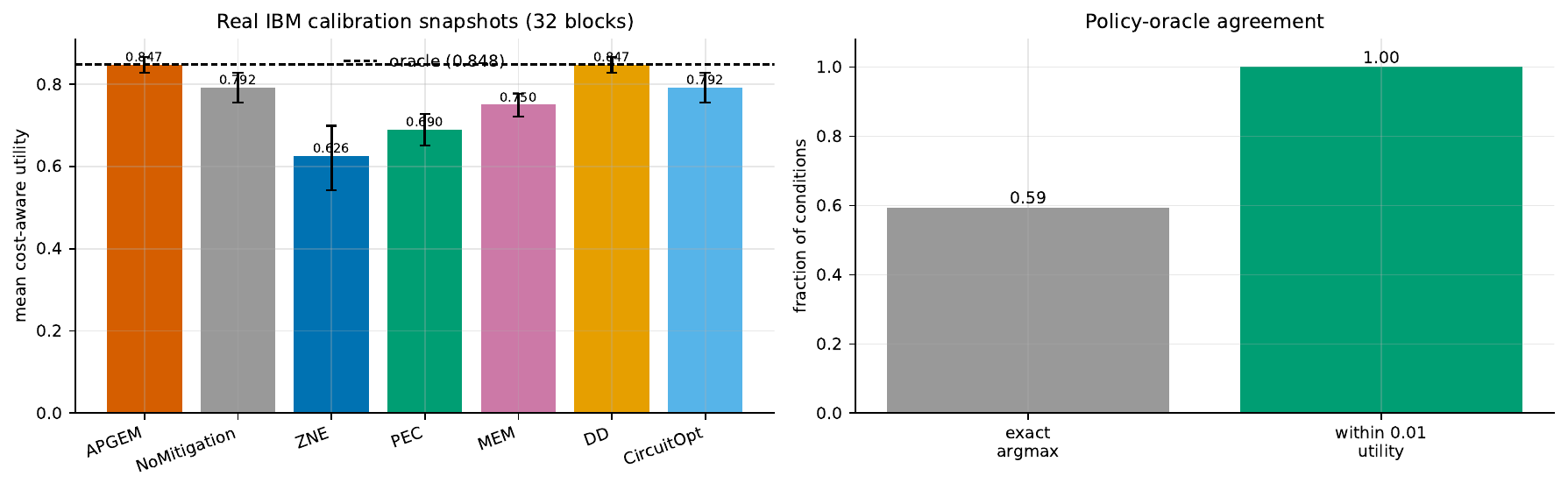}
\caption{Validation on real IBM calibration snapshots. Left: mean utility per strategy over $32$ device-condition blocks. Right: exact-argmax versus within-tolerance agreement with the oracle.}
\label{fig:calibrated}
\end{figure}

Everything so far uses constructed noise grids. To see whether these findings carry over to real hardware, we replay the controller on $32$ device-condition blocks built from eight IBM calibration snapshots; the measured parameter ranges for those snapshots are given in Table~\ref{tab:calenv}. On these snapshots the controller attains $99.9\%$ of the expected-utility oracle (Friedman $\chi^{2}=128.68$, $p=2.4\times10^{-25}$ over the $32$ blocks), and every single selection falls within $0.01$ utility of the oracle: exact argmax match is only $59.4\%$, but within-tolerance agreement is a full $100\%$ (Table~\ref{tab:cal}, Fig.~\ref{fig:calibrated}). The same regime structure we saw in the simulated grid shows up here too. Dynamical decoupling becomes competitive on the coherent-error snapshots, where its average rank ($2.33$) is statistically indistinguishable from APGEM's ($2.27$, Wilcoxon $p=0.32$), and readout mitigation is correctly passed over, since the measured readout errors ($0.014$ to $0.037$) sit well below the threshold at which its two-fold sampling cost would be worth paying.

\begin{table}[htbp]
\centering
\caption{Range of measured parameters across the eight IBM calibration snapshots used for validation.}
\label{tab:calenv}
\begin{tabular}{lcccc}
\toprule
 & Readout error & Two-qubit error & $T_1$ ($\mu$s) & $T_2$ ($\mu$s) \\
\midrule
Range across 8 devices & 0.014--0.037 & 0.006--0.013 & 62--148 & 39--115 \\
\bottomrule
\end{tabular}
\end{table}

\begin{table}[htbp]
\centering
\caption{Validation summary on the $32$ real-device blocks. The controller nearly saturates the oracle, and all disagreements are within-tolerance near-ties.}
\label{tab:cal}
\begin{tabular}{lc}
\toprule
Quantity & Value \\
\midrule
Blocks                     & 32 \\
APGEM utility (\% of oracle) & 99.9\% \\
Exact argmax match         & 0.594 \\
Within-tolerance agreement ($\leq 0.01$) & 1.000 \\
Friedman $\chi^{2}$        & 128.68 \\
Friedman $p$               & $2.4\times10^{-25}$ \\
\bottomrule
\end{tabular}
\end{table}

\section{Discussion}
\label{sec:discussion}
The main thing these results say is that no single mitigation technique is right often enough to justify committing to it. PEC wins on the deterministic and generative paradigms, ZNE takes over at low noise, and plain NoMitigation wins outright once the noise gets heavy enough that the extra shots stop being worth it (Table~\ref{tab:regime}, Fig.~\ref{fig:policymap}). Any fixed rule pays for this somewhere: an always-off policy leaves quality on the table in the moderate-noise band where PEC would help, and an always-on PEC policy would spend a $4\times$ shot budget in regimes where that quality gain barely registers (Table~\ref{tab:tech}). What the controller buys is the ability to resolve that trade-off per block instead of picking one answer in advance, and it does not seem to matter much which contextual bandit does the resolving, LinUCB and Thompson sampling come out within half a point of each other and both clear the context-free baseline by a wide margin (Section~\ref{sec:r-ctrl}).

What makes this more than a curve-fitting exercise is that the same controller, given no signal about which paradigm it is looking at, still settles on five different policies for the five paradigms (Section~\ref{sec:r-para}), and its decision boundaries line up with physically sensible quantities -- two-qubit error, readout error, circuit depth rather than anything specific to the training grid (Fig.~\ref{fig:selcontext}). The clearest evidence of this, though, is the real-device check in Section~\ref{sec:r-cal}: on eight IBM calibration snapshots that the controller never trained on and that carry a different noise profile than the constructed grid, it still recovers $99.9\%$ of oracle utility, and dynamical decoupling correctly comes back into favour exactly on the snapshots where coherent errors dominate. A controller that had simply overfit the training grid would have no reason to reproduce that regime-specific behaviour on data it has never seen.

It is also worth being honest about what argmax match does and does not tell you here. Only $57.5\%$ of blocks see an exact match with the oracle's choice, which sounds mediocre until you notice the controller still captures $95.5\%$ of oracle utility. The regret decomposition in Fig.~\ref{fig:regret} explains the gap: roughly two-thirds of what looks like error is really the irreducible noise floor of comparing against a per-realization ex-post oracle, not a mistake the controller could have avoided. QRL is the clearest case, its stochastic routing pushes that noise floor to $0.1116$, almost two orders of magnitude above VQE's $0.0006$ (Table~\ref{tab:regret}), and it is also the paradigm with the lowest argmax match, because its blocks are simply the hardest to score, not because the controller understands them worst; it still reaches $94\%$ of QRL's own oracle utility. The real-device data show the same thing from another angle: every selection lands within $0.01$ utility of the oracle even though exact argmax match is only $59.4\%$ (Table~\ref{tab:cal}). Any time the environment has real sampling variance, argmax match alone will make a controller that is doing nearly everything right look considerably worse than it is.

None of this works without folding cost into the objective in the first place. PEC's quality lead over NoMitigation is genuine, but it comes at $4\times$ the shots, and ZNE's smaller lead costs $3\times$ (Table~\ref{tab:tech}); an objective built on raw quality alone would push the controller toward PEC almost everywhere, and Section~\ref{sec:r-noise} shows that would be the wrong call once the noise gets high enough that NoMitigation is nearly as good for a fraction of the price. The scalability sweep matters for a related reason: because the per-decision shot budget is fixed regardless of qubit count, running the controller itself does not get more expensive as circuits grow from three to eight qubits, even as the ceiling on achievable quality drops with circuit size (Section~\ref{sec:r-scal}).

A few things limit how far we would push these conclusions. The $400$-interaction horizon used for the main results sits on the near-flat part of the utility curve, but regret is still falling at $1000$ interactions (Table~\ref{tab:horizon}), so a deployment willing to spend more shots on learning could plausibly do a bit better than what we report. The real-device check covers eight calibration snapshots and $32$ blocks, which is enough to reproduce the regime structure seen in simulation but still a narrower slice of hardware conditions than the $80$-block constructed grid, and we would not claim the same margins hold across every noise profile a fleet of devices might present. The six mitigation arms and five paradigms here were chosen to be representative rather than exhaustive; extending the arm set to techniques that can be composed rather than chosen exclusively is an obvious next step. And everything the controller does is downstream of the cost-aware utility defined in Section~\ref{exp} -- a practitioner with a different quality-cost balance, or a hard latency budget instead of a soft shot penalty, would need to re-derive that utility before expecting the same controller to make the same calls.

\section{Conclusion}
\label{sec:conclusion}
The question we set out to answer was a practical one: given a menu of error-mitigation techniques, none of which is uniformly best, can a controller that only sees a block's noise context and a shared feedback signal do better than committing to any single one of them in advance? On the evidence presented here, the answer is yes, and by a margin large enough to matter. Across $80$ device-condition blocks spanning five learning paradigms and six mitigation arms, the contextual controller recovers $95.5\%$ of oracle utility while cutting expected regret nearly in half relative to the best static policy, and the gain survives a Friedman test, post hoc Nemenyi comparisons, and paradigm-by-paradigm Wilcoxon tests against every one of the six fixed arms. It is not a fragile result tied to one algorithm: swapping LinUCB for Thompson sampling changes the outcome by less than half a percentage point, which is a reasonable basis for attributing the gain to the act of conditioning on context rather than to any particular exploration strategy. What we find more interesting than the headline number is what the controller does with the freedom it is given. It does not converge to a single favourite arm and hide behind it; it recovers five distinct paradigm-specific policies from a shared interface that never tells it which paradigm it is looking at, and its decision boundaries track physically interpretable quantities -- two-qubit error, readout error, circuit depth -- rather than artefacts of how the training grid was constructed. The regret decomposition adds an important qualification to the headline numbers: a large share of what looks like disagreement with the oracle is in fact irreducible sampling noise rather than a correctable error, which matters for how this kind of result should be reported and interpreted more generally, not just in this study. And the fact that the same policy structure re-emerges, largely unchanged, on real IBM calibration snapshots the controller never trained on is, to us, the result that carries the most weight. Simulated noise grids are convenient but they are also generous; agreement between simulation and hardware is the test that a method either passes or does not, and this one passes it. None of this argues that contextual selection is a solved problem. The arm set studied here is representative rather than exhaustive, the real-device validation is smaller than the constructed grid it is meant to corroborate, and everything the controller optimizes for is downstream of one particular cost-aware utility function that a different deployment might reasonably define differently. But the broader claim we think this work supports is a modest and, we hope, durable one: as mitigation techniques continue to multiply and their relative value keeps shifting with hardware, workload, and noise regime, treating mitigation choice as a policy to be learned, rather than a setting to be fixed, is a cheap change that gives back a large and consistent fraction of what an omniscient oracle would achieve. Extending this idea to mitigation techniques that compose rather than compete, and testing it across a wider population of real devices, are the two directions we see as most likely to determine whether that claim continues to hold as the field's toolbox grows.

\bibliographystyle{unsrt}
\bibliography{sample}
\end{document}